\documentclass[aps,prd,preprint,a4paper,showpacs,nofootinbib,superscriptaddress]{revtex4-2}
\usepackage{bm}
\usepackage{indentfirst}
\usepackage{amsmath}
\usepackage{graphicx}
\usepackage{amssymb}
\usepackage{subfigure}
\usepackage{amssymb}
\usepackage{hyperref}
\usepackage{epstopdf}
\usepackage[section]{placeins}

\usepackage[utf8]{inputenc}
\hypersetup{
    colorlinks=true,
    linkcolor=red,
    citecolor=blue,
}
\usepackage{color}
\usepackage[T1]{fontenc}
\usepackage{txfonts}
\usepackage{orcidlink}

\begin{document}


\title{Phase structure of holographic superconductors in an Einstein-scalar-Gauss-Bonnet theory with spontaneous scalarization}

\author{Hong Guo}
\email{gravhguo@gmail.com}
\affiliation{Escola de Engenharia de Lorena, Universidade de São Paulo, 12602-810, Lorena, SP, Brazil}

\author{Wei-Liang Qian}
\email{wlqian@usp.br (corresponding author)}
\affiliation{Escola de Engenharia de Lorena, Universidade de São Paulo, 12602-810, Lorena, SP, Brazil}
\affiliation{Faculdade de Engenharia de Guaratinguet\'a, Universidade Estadual Paulista, 12516-410, Guaratinguet\'a, SP, Brazil}
\affiliation{Center for Gravitation and Cosmology, College of Physical Science and Technology, Yangzhou University, 225009, Yangzhou, China}

\author{Bean Wang}
\affiliation{Department of Physical Sciences and Applied Mathematics, Vanguard University, Costa Mesa, CA 92626, USA}

\begin{abstract}
Holographic superconductor phase transition and spontaneous scalarization are triggered by the instability of the underlying vacuum black hole spacetime.
Although both hairy black hole solutions are closely associated with the tachyonic instability of the scalar degree of freedom, they are understood to be driven by distinct causes.
It is, therefore, interesting to explore the interplay between the two phenomena in the context of a scenario where both mechanisms are present.
To this end, we investigate the Einstein-scalar-Gauss-Bonnet theory in asymptotically anti-de Sitter spacetime with the presence of a Maxwell field. 
Even though different origins for the tachyonic mass behave independently and can be recognized by the distinctive natures of their effective potentials, it is shown that near the transition curve, the holographic superconductor and spontaneous scalarization are found to be largely indistinguishable.
This raises the question of whether the hairy black holes triggered by different mechanisms are smoothly joined by a phase transition or whether these are actually identical solutions.
To assess the transition more closely, we evaluate the phase diagram in terms of temperature and chemical potential and discover a smooth but first-order transition between the two hairy solutions by explicitly evaluating Gibbs free energy and its derivatives.
In particular, one can elaborate a thermodynamic process through which a superconducting black hole transits into a scalarized one by raising or decreasing the temperature. 
Exhausting the underlying phase space, we analyze the properties and the interplay between the two hairy solutions.

\end{abstract}

\maketitle

\newpage

\section{Introduction}\label{sec=intro}

As an enigmatic prediction of General Relativity, the black hole is an extreme manifestation of the spacetime curvature. 
In the past millennium, owing to the continuous endeavor in astrophysics regarding observation associated with both electromagnetic and gravitational wave channels, the black hole is arguably the most notable astrophysical object~\cite{book-blackhole-Frolov, Berti:2015itd}.  
Specifically, recent decades have witnessed unprecedented advances in gravitational wave detection, achieved by LIGO-Virgo Collaboration, which involves more than one hundred black hole binary mergers~\cite{LIGOScientific:2016vlm, LIGOScientific:2018mvr, LIGOScientific:2019fpa, LIGOScientific:2021djp}. 
These prominent astrophysical events provide a crucial means to investigate extreme gravitational phenomena. 
In particular, the most complex dynamics and extreme gravitational conditions occur in the vicinity of the black hole's horizon. 
As it plays a pivotal role by connecting theoretical speculations with astrophysical observations, the relevant studies have triggered much attention in the literature~\cite{Barack:2018yly, Berry:2019wgg, Babak:2017tow}.

One pertinent topic in black hole physics concerns a series of ``no-hair'' theorems and their evasion~\cite{Ruffini:1971bza, Chrusciel:2012jk, Carter:1971zc}.
These theorems state that all the information about a black hole is determined by its mass, charge, and angular momentum. 
On the other hand, substantial insights can also obtained by exploring the scenarios when the prior condition of such theorems becomes invalid.
The latter might give rise to hairy black hole solutions owing to various mechanisms~\cite {Herdeiro:2015waa, Sotiriou:2015pka, Herdeiro:2014goa}. 
The celebrated holographic superconductor is primarily owing to the tachyonic instability in the asymptotically anti-de Sitter (AdS) spacetime, complemented with the presence of a Maxwell field and a charged scalar~\cite{Gubser:2008px, Hartnoll:2008kx}.
In this framework, the asymptotic AdS spacetime is crucial to evading the prerequisite of no-hair theorems in asymptotically flat spacetimes~\cite{Sudarsky:2002mk}.

More recently, an alternative mechanism for hairy black holes, known as black hole spontaneous scalarization, has been proposed.
In its original form, it refers to the scenario where the scalar degree of freedom is nonminimally coupled to the Gauss-Bonnet curvature in an asymptotically Minkowski spacetime, giving rise to the emergence of hairy black hole solutions~\cite{Doneva:2017bvd, Silva:2017uqg, Antoniou:2017acq}.
In this regard, one pivotal feature of the theory is its ability to evade the no-hair theorem in asymptotically flat spacetimes. 
In the literature, the notion of black hole spontaneous scalarization quickly garnered significant attention.
It was observed that black hole spontaneous scalarization may take place in a more general context where the scalar field is nonminimallly coupled to source terms furnished by various types of matter fields, inclusively the Maxwell invariant~\cite{Herdeiro:2018wub}, Chern-Simons invariant~\cite{Brihaye:2018bgc}, and the Ricci scalar~\cite{Herdeiro:2019yjy}. 
Further developments on spontaneous scalarization involve spin-induced~\cite{Dima:2020yac,Herdeiro:2020wei,Berti:2020kgk,Doneva:2023oww}, nonlinear~\cite{Doneva:2021tvn,Lai:2023gwe,Doneva:2022yqu,Pombo:2023lxg}, and dynamical descalarization scenarios~\cite{Silva:2020omi,Kuan:2021lol,Zhang:2021etr,Zhang:2021ybj,Zhang:2021nnn,Liu:2022fxy}. 
This mechanism is also extendable to cases with a cosmological constant~\cite{Bakopoulos:2018nui,Brihaye:2019gla,Brihaye:2019dck,Guo:2021zed,Promsiri:2023yda,Zou:2023inv,Marrani:2022hva}, contributing to its prominence.
A comprehensive survey of recent progress can be found in the review~\cite{Doneva:2022ewd}.

Notably, spontaneous scalarization is also attributed to the instability of the underlying hair-free black hole solution.
The latter is demonstrated as an effective tachyonic mass in the master equation governing the linearized scalar perturbations.
In particular, the stability of the corresponding ``bald'' black hole solution can be analyzed by explicitly evaluating the quasinormal frequencies~\cite{Doneva:2017bvd, Blazquez-Salcedo:2018jnn, Myung:2018iyq}. 
In Refs.~\cite{Blazquez-Salcedo:2018jnn, Myung:2018iyq}, the onset of spontaneous scalarization is recognized as when the purely imaginary quasinormal modes touch the origin. 
More specifically, bound state solutions of the scalar field were derived in~\cite{Silva:2017uqg, Silva:2018qhn, Antoniou:2021zoy}, for which the occurrence of spontaneous scalarization is shown to coincide with that for the marginally stable quasinormal mode encountered in~\cite{Blazquez-Salcedo:2018jnn, Myung:2018iyq}. 
A sufficient condition for the tachyonic instability is attained when the effective potential for the scalar perturbations $V_\mathrm{eff}$ can be essentially viewed as a potential well, namely, $\int^{+\infty}_{-\infty} dr_* V_\mathrm{eff}(r_*)<0$.
This condition guarantees~\cite{buell1995potentials} at least one bound state, indicating the instability of a bald black hole, which typically occurs as the coupling exceeds a critical value. 
It is noteworthy to point out that the occurrence of superluminal propagation in the system with a substantial Gauss-Bonnet term is known in the literature and has been explored in the context of the AdS/CFT correspondence~\cite{Brigante:2008gz,Brigante:2007nu,Buchel:2009tt,Hofman:2008ar,Hofman:2009ug,Camanho:2009vw}. 
In the present scenario, such an effect becomes ``dynamic'' as it is nonlinearly coupled to a scalar field. 
Subsequently, rather than a ``static'' bound for metric parameters, one acquires a physical instability associated with the scalar degree of freedom that eventually gives rise to a scalarized hairy black hole.

As discussed in~\cite{Blazquez-Salcedo:2018jnn}, it is also essential to note that the tachyonic instability, being a sufficient condition, does not always align with the onset of marginally stable quasinormal modes.
Nonetheless, the instability of the bald black hole solution is ascertained by the unstable quasinormal modes, while the subsequent transition to a hairy black hole can be confirmed by explicitly deriving the non-vanishing bound state solution~\cite{Silva:2017uqg, Silva:2018qhn, Antoniou:2021zoy} and evaluating the entropies of the scalarized black hole and compare it against that of its bald counterpart~\cite{Doneva:2017bvd}.
Given that the occurrence of unstable quasinormal modes is a weaker condition whose onset does not always warrant tachyonic instability, it was argued in~\cite{Myung:2018iyq} that spontaneous scalarization is caused by the Gregory-Laflamme instability~\cite{Gregory:1993vy}. 
In other words, another mechanism might trigger scalarization before the effective potential eventually becomes a potential well.

The present study is motivated by the above consideration to explore further the properties and relation between the two mechanisms for hairy black holes.
To this end, we employ the Einstein-scalar-Gauss-Bonnet theory in asymptotically AdS spacetime, a scenario where both relevant mechanisms are present.
Specifically, a charged scalar field is coupled to the Gauss-Bonnet invariant in such a framework.
On the one hand, the minimal coupling between the scalar field and the Maxwell field leads to a tachyonic instability, forming a $s$-wave holographic superconductor~\cite{Hartnoll:2008vx}. 
On the other hand, the scalar field is coupled to the Gauss-Bonnet curvature, giving rise to spontaneous scalarization~\cite{Doneva:2017bvd}.
Although both instabilities imply a transition to the hairy black hole, it is not entirely clear whether the hairy black hole are equivalent, given that the resulting profiles of the fields are largely indistinguishable near the transition point.
We scrutinize this point by evaluating the phase diagram in terms of temperature and chemical potential, and explicitly calculating the Gibbs free energy and its derivatives.
In particular, we identify a rather smooth but first-order transition between the two hairy solutions.
By exhausting the parameter space of the underlying black hole metric, we analyze the properties and the interplay between the two hairy solutions.
The holographic superconductor phase is found to flip over to the other side of the transition curve when the temperature drops below the critical value corresponding to vanishing Gauss-Bonnet coupling.
Moreover, it is pointed out that the two mechanism can also be distinguished by their specific shapes of the effective potentials. 

The remainder of the paper is organized as follows. 
In the following section, we elaborate on the Einstein-scalar-Gauss-Bonnet model, the relevant equations of motion, and the corresponding boundary conditions. 
The numerical scheme is presented in Sec.~\ref{sec=phase}, which is then used to derive the hairy black hole solutions and subsequently the phase diagram.
We explore the properties of the obtained solutions associated with the holographic superconductor and spontaneous scalarization.
Furthermore, the phase diagram of the model is presented in terms of temperature and chemical potential.
We analyze the specific shapes of the effective potentials reflecting the underlying instabilities of the underlying gravitational system. 
The last section is devoted to further discussions and concluding remarks.

\section{Einstein-scalar-Gauss-Bonnet Model}\label{sec=model}

In this section, we elaborate on the Einstein-scalar-Gauss-Bonnet model employed in the present study.
The action consists of a charged massive scalar field $\psi$ nonminimally coupled to the Gauss-Bonnet invariant with the presence of a Maxwell field in an asymptotically AdS spacetime~\cite{Guo:2020sdu, Guo:2020zqm}
\begin{equation}\label{eq=action}
    S=\frac{1}{16 \pi G_N} \int \, d^4 x \, \sqrt{-g} \left[R+\frac{6}{L^2}-\frac{1}{4}F_{\mu\nu}F^{\mu \nu}-|D_\mu \psi|^2-m^2 |\psi|^2 +f(\psi) \mathcal{R}^2_{GB}\right]~,
\end{equation}
where the Gauss-Bonnet curvature $\mathcal{R}^2_{GB}=R^2-4R_{\mu\nu}R^{\mu\nu}+R_{\mu\nu\alpha\beta}R^{\mu\nu\alpha\beta}$, $f(\psi)$ describes the non-minimal coupling between the scalar and the spacetime curvature. 
The covariant derivative is defined by $D_\mu=\nabla_\mu-iqA_\mu$, where the scalar's electric charge $q$ measures its coupling to the Maxwell field. 
Also, $G_N$ is Newton's constant, and $L$ represents the curvature radius of the AdS spacetime. 

On the one hand, as the charge $q$ vanishes, the action falls back to a more straightforward case that furnishes spontaneous scalarization~\cite{Doneva:2017bvd}. 
On the other hand, if one assumes $f(\psi)=0$, the model is essentially a s-wave holographic superconductor~\cite{Hartnoll:2008vx}. 
We note that to guarantee that the coupling function $f(\psi)$ can induce the spontaneous scalarization, the following specific form is adopted~\cite{Doneva:2017bvd}
\begin{equation}\label{eq=sca_coup}
    f(\psi) =\frac{\lambda^2}{2}\left( 1- e^{-\psi^2}\right)~.
\end{equation}
where the strength $\lambda$ is a constant, so that $f'(0)=0$ and $f''(0)>0$~\cite{Doneva:2017bvd}.

In the asymptotically AdS spacetime, we consider the following metric ansatz
\begin{align}\label{eq=metric}
    ds^2=-g(r) dt^2+\frac{1}{g(r)}dr^2+r^2\left(dx^2+dy^2\right)~,
\end{align}
where 
\begin{equation}\label{eq=backgroundg}
    g(r)=\frac{r^2}{L^2}-\frac{M}{r}~,
\end{equation}
and the Hawking temperature reads 
\begin{equation}
    T=\frac{g'(r_h)}{4\pi}.
\end{equation}
In the probe limit, by varying the action Eq.~\eqref{eq=action} w.r.t. the scalar and electromagnetic degrees of freedom, one finds the following equations of motion 
\begin{align}
    &\nabla_\mu \nabla^\mu \psi -\left(m^2+q^2A_\mu A^\mu \right)\psi+\frac{1}{2}f'(\psi) \mathcal{R}^2_{GB}=0~, \label{eq=scalar}\\
    &\nabla_\alpha F^{\alpha\mu}=2q^2A^\mu \psi^2~.\label{eq=maxwell}
\end{align}
By considering spherically symmetric case where $A_\mu dx^\mu=\phi(r)dt,\ \psi=\psi(r)$, the equations are further simplified to read
\begin{align}
    &\phi''(r)+\frac{2}{r}\phi'(r)-\frac{2q^2\psi(r)^2}{g(r)}\phi(r) = 0 \label{eq=phi}~,\\
    &\psi''(r)+\left(\frac{2}{r}+\frac{g'(r)}{g(r)}\right)\psi'(r)+ \frac{q^2 \phi(r)^2-m^2g(r)}{g(r)^2}\psi(r)+\frac{{\cal R}^2_{GB}}{2g(r)}f'(\psi)= 0~,\label{eq=psi}
\end{align}
where the Gauss-Bonnet curvature is evaluated as $\mathcal{R}^2_{GB}=\frac{4}{r^2}\left[g'(r)^2+g(r)g''(r)\right]$.

Using the tortoise coordinate $dr_*=\frac{dr}{g(r)}$ and denoting $\psi=\frac{\varphi}{r}$, the Klein-Gordon equation \eqref{eq=psi} can be brought into a Schrodinger-like form as
\begin{equation}\label{eq=schrodinger}
    \frac{\partial^2 \varphi(r)}{\partial r_{*}^2}-V_\mathrm{eff}(r) \varphi(r)=0~,
\end{equation}
where the effective potential of the scalar field is 
\begin{equation}\label{eq=potential}
    V_\mathrm{eff}(r)=g(r)\left(\frac{g^{\prime}(r)}{r}+m^2-\frac{q^2}{g(r)}\phi(r)^2-\frac{\lambda^2}{2} \mathcal{R}_{G B}^2\right).
\end{equation}
To derive Eqs.~\eqref{eq=schrodinger} and~\eqref{eq=potential}, it is noted that the scalar and electromagnetic fields are treated as perturbations and in particular, we have only kept the leading term in the expansion $\frac{df(\psi)}{d\psi}\simeq\psi\bigg(1-\psi^2+\frac{1}{2}\psi^4+\mathcal{O}(\psi^6)\bigg)$.

We proceed to discuss the boundary conditions for the above equations of motion. 
It is important to note that the boundary conditions are derived based on generic requirements for the fields to be regular and the asymptotical forms of the equations of motion.
These conditions are universal for distinct hairy black hole solutions associated with different physical natures.

At the event horizon, $r=r_h$, the scalar and Maxwell field must be regular.
From Eq.~\eqref{eq=phi}, the last term on the l.h.s. indicates 
\begin{equation}
\phi(r_h)=0 .\label{eq=phi-horizon}
\end{equation}
By substituting it into Eq.~\eqref{eq=psi}, we have
\begin{equation}\label{eq=psi-horizon}
    \psi'(r_h)=\dfrac{L^2}{3r_h} \left(m^2-\dfrac{18 \lambda^2 e^{-\psi(r_h)^2}}{L^4}\right)\psi(r_h)~.
\end{equation}
At spatial infinity, by analyzing the leading contributions, the asymptotical behaviors of scalar and Maxwell fields are found to be
\begin{align}
    \phi(r) &=\mu-\frac{\rho}{r}~, \label{eq=phi-inf} \\
    \psi(r) &= \frac{\psi_1}{r^{\Delta_-}}+\frac{\psi_2}{r^{\Delta_+}}~, \label{eq=psi-inf}
\end{align}
where $\Delta_{\pm}=\frac{3\pm\sqrt{9+4m^2_eL^2}}{2}$ and the effective scalar mass is defined as $m^2_e=m^2-12\frac{\lambda^2}{L^4}$.
One enforces the condition 
\begin{equation}
\psi_2=0 ,\label{bd-condition}
\end{equation}
so that the condensation is turned on without a source on the AdS boundary.

In the framework of the holographic principle, $\mu$ is the chemical potential, and $\rho$ describes the charge density of the field theory. 
The condensate of the scalar operator $O$ in the field theory dual to the field $\psi$ is given by
\begin{equation}
    <O_1>=\sqrt{2}\psi_1.
\end{equation}
The asymptotical behavior of Eq.~\eqref{eq=phi-inf} can be used to extract the density and chemical potential. 

It is not difficult to observe that there is a scaling symmetry in the model; namely, a hairy black hole solution continues to be valid under the following scaling transform
\begin{equation}
r\to a r, \ \ (t, x, y,) \to  (t, x, y)/a, \label{scale1}
\end{equation}
accompanied by
\begin{equation}
M \to a^3 M, \ \ L\to L,\ \ r_h\to a r_h,\ \ g \to a^2 g ,\label{scale2}
\end{equation}
and
\begin{align}
\phi(r) \to a\phi(r/a), \ \ \psi(r)\to \psi(r/a),\ \ \mathcal{R}_{GB}\to \mathcal{R}_{GB}, \label{scale3}\\
q\to q, \ \ m\to m,\ \ \lambda\to \lambda .\label{scale4}
\end{align}
In this regard, we redefine temperature and chemical potential in a scaling transform invariant fashion, namely, $\tilde{T}=\frac{T}{T_c}$ and $\tilde{\mu}=\frac{\mu}{T_c}$, where $T_c$ is the transition temperature at vanishing Gauss-Bonnet coupling.

\section{Two types of hairy black hole solutions}\label{sec=phase}

\begin{figure}[thbp]
    \centering
    \includegraphics[width=0.48\linewidth]{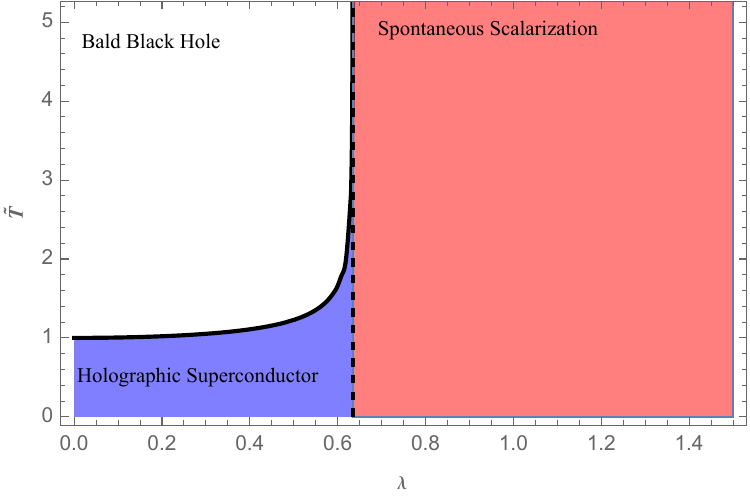}
    \caption{The phase diagram shown in terms of temperature $\tilde{T}$ and Gauss-Bonnet coupling $\lambda$. 
    The blue region represents the holographic superconductor phase, and the solid black curve indicates the boundary where the transition to a bald black hole occurs.
    The red region corresponds to the spontaneous scalarization phase, so the black dashed line represents the critical value of the Gauss-Bonnet coupling.
    The calculations have been carried out by adopting $r_h=L=1$, $q=1$ and $m_e^2=-2$.}
    \label{fig=phase-diagram}
\end{figure}

\begin{figure}[thbp]
    \centering
    \includegraphics[width=0.325\linewidth]{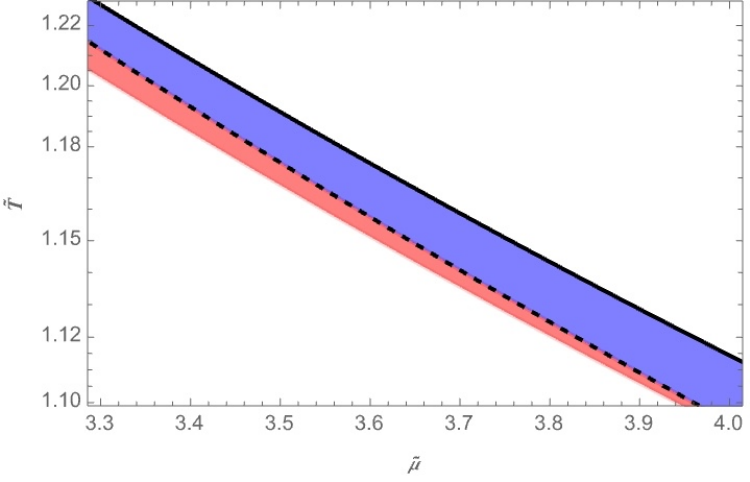}
    \includegraphics[width=0.325\linewidth]{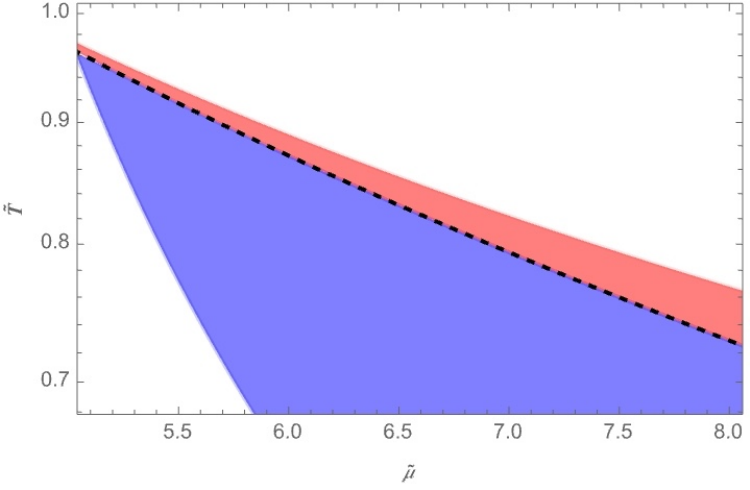}
    \includegraphics[width=0.325\linewidth]{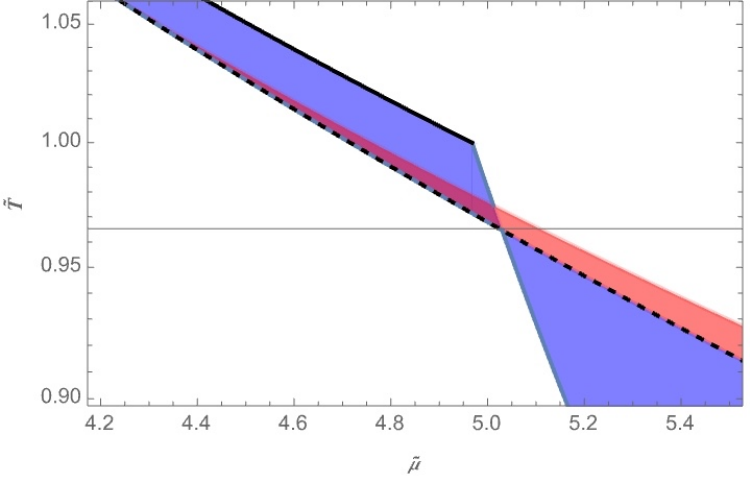}
    \caption{The phase diagram shown in terms of temperature $\tilde{T}$ and chemical potential $\tilde{\mu}$.
    The conventions introduced in Fig.~\ref{fig=phase-diagram} have been adopted, and the calculations have been carried out using the same parameters.
    The left panel shows the phase diagram in the high temperature and small chemical potential region.
    The middle panel corresponds to the low temperature and large chemical potential region.
    The right panel shows the region where the holographic superconductor phase flips to the other side of the spontaneous scalarization one.}
    \label{fig=phase-diagram2}
\end{figure}

In this section, we demonstrate that two distinct types of hairy black hole solutions associated with, respectively, holographic superconductor and scalarization, coexist in the model.
Using the numerical scheme presented in Sec.~\ref{sec=NumScheme}, the phase diagram will be evaluated and presented in terms of temperature and Gauss-Bonnet coupling shown in Fig.~\ref{fig=phase-diagram}.
The system comprises three phases: thermalized vacuum represented by a phase of bald black hole solutions, holographic superconductor, and scalarization phases indicated by hairy black hole solutions.
As will be validated in Sec.~\ref{sec=HBH_transition}, to identify and explore the transition between the two hairy black hole phases, one may also present the phase diagram in terms of temperature and chemical potential, shown in Fig.~\ref{fig=phase-diagram2}.
We note that the latter two thermodynamic quantities are independent ones not constrained by scaling laws, while they facilitate the study of phase transition's order.
Moreover, the thermodynamic properties of their interpretation in the dual field theory are then elaborated, and in particular, the free energy of obtained spacetime configurations are evaluated.
We relegate the stability analysis of the underlying bald black hole solutions to Sec.~\ref{sec=instab}.

\subsection{Numerical procedure}\label{sec=NumScheme}

\begin{figure}[thbp]
    \centering
    \includegraphics[width=0.48\linewidth]{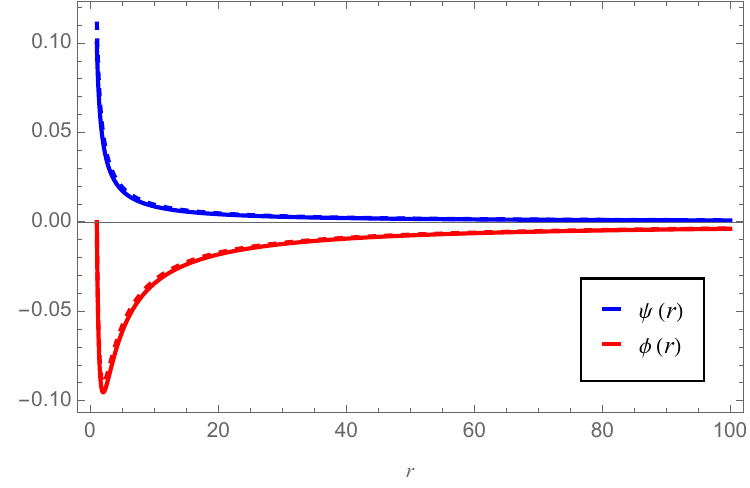}
    \caption{Radial profiles of the scalar(blue) and Maxwell fields(red). The solid lines represent the holographc superconductor solution with $\lambda=0.63,\psi(r_h)=0.1,\phi'(r_h)=0.38$, and the dashed lines are the spontaneous scalarization solution with $\lambda=0.64,\psi(r_h)=0.112,\phi'(r_h)=0.36$.}
    \label{fig=solution}
\end{figure}

The hairy black hole solutions are derived numerically using the shooting method.
The scalar and Maxwell fields are evaluated by the equations of motion Eqs.~\eqref{eq=phi} and ~\eqref{eq=psi} using numerical integration.
As mentioned above, the two types of hairy black holes are encountered by adopting the same boundary conditions discussed in the last section.
Besides, from a mathematical perspective, although they originated from different physical mechanisms and reside in different regions of the parameter space, the algorithm to derive these solutions is mainly identical, specified as follows.
One starts the numerical integration at the horizon $r=r_h$ where the two lowest Taylor expansion coefficients are determined by assuming the values of $\psi(r_h)$ and $\phi'(r_h)$ and using Eqs.~\eqref{eq=phi-horizon} and~\eqref{eq=psi-horizon}.
The shooting procedure is accomplished by enforcing the condition Eq.~\eqref{bd-condition} at the boundary.
As for the one-dimensional Schrodinger-like equation, the number of nodes corresponds to the energy level.
In the present study, we will focus on the ground state by only considering the solution without any node, as shown in Fig.~\ref{fig=solution}.

For simplicity, the numerical calculations are carried out using $r_h=L=1$ and $q=1$.
In particular, following~\cite{Hartnoll:2008vx,Hartnoll:2008kx,Franco:2009yz,Bao:2021wfu,Pan:2021jii}, we set the effective mass to $m^2_e=-2$, which is above the Breitenlohner-Freedman bound for stability.
By comparing the obtained radial profiles $\psi(r)$ and $\phi(r)$ to the asymptotical form Eq.\eqref{eq=psi-inf}, the values of $\mu$ and $\rho$ are extracted.
The resulting family of solutions is characterized by two variables, namely, the coupling $\lambda$ and $\psi(r_h)$. 
One then employs the system's scaling invariance by using Eqs.~\eqref{scale1}-\eqref{scale4} to cover the remainder of the parameter space.

When the Gauss-Bonnet coupling $\lambda$ is sufficiently tiny, specifically $\lambda<\lambda_c\approx 0.6339$, it is shown that a holographic superconductor phase is encountered featuring a transition at specific value of $\phi'(r_h)$~\cite{Guo:2020sdu}, which can be effectively viewed to occur at a specific temperature employing the scaling Eqs.~\eqref{scale1}-\eqref{scale4}.
In particular, it would be the only relevant mechanism in the present model to form a hairy black hole if one takes the limit $\lambda \to 0$.
On the other hand, hairy solutions due to spontaneous scalarization can be obtained by employing the same numerical scheme at larger Gauss-Bonnet coupling.
Unlike a holographic superconductor, such a solution persists even when the charge of the scalar vanishes.

To explore the phase structure of the system, one may first derive the above solutions in the specific region of the phase space and then continuously vary the parameters to the region of interest where both mechanisms are potentially relevant.
Starting from a hairy black hole solution associated with a holographic superconductor, one explores the solution space by continuously increasing the Gauss-Bonnet coupling.
As demonstrated in Fig.~\ref{fig=phase-diagram}, it turns out that the transition at the critical temperature diverges as $\lambda\to \lambda_c$ from below.
Both mechanisms persist for finite charge $q$ and coupling $\lambda$.

As an illustration, the radial profiles of the scalar and Maxwell fields are shown in Fig.~\ref{fig=solution} in blue and red curves.
The solid curves represent the numerical results for a holographically superconducting black hole with the metric parameters $\psi(r_h)=0.1, \phi'(r_h)=0.38$, and $\lambda=0.63$. 
The dashed lines show the radial profiles of a spontaneous scalarization black hole with the parameters $\psi(r_h)=0.112, \phi'(r_h)=0.36$, and $\lambda=0.64$. 
It is observed that the condensation of the scalar field while the temporal component of the electromagnetic field vanishes at the horizon.
As the fundamental states, the obtained radial profiles do not contain any nodes. 
Notably, the two distinct types of hairy black holes somehow bear a strong resemblance if they are near the critical coupling $\lambda_c$.
As the fields' radial profiles of the two cases are largely indistinguishable, it is not entirely clear whether the hairy black hole solutions residing on the two sides of $\lambda= \lambda_c$ are different ones and, even if they are potentially triggered by different mechanisms.
In this regard, it is interesting to analyze the properties of the two phases and the transition between them further in the context of the system's phase structure.

\begin{figure}[thbp]
    \centering
    \includegraphics[width=0.48\linewidth]{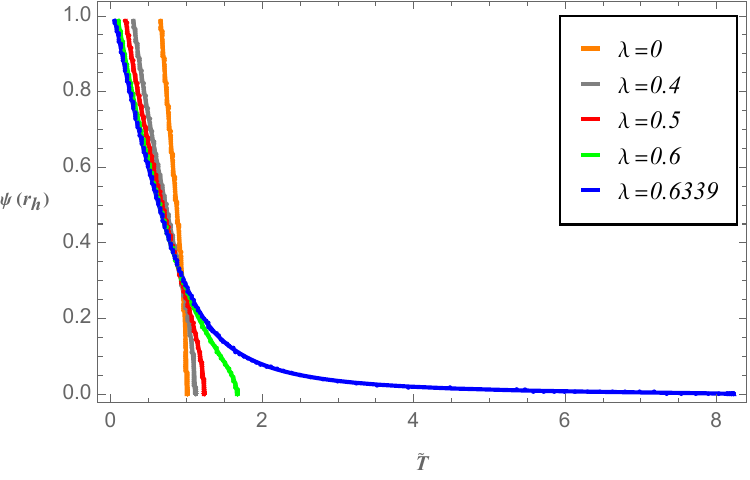}
    \includegraphics[width=0.48\linewidth]{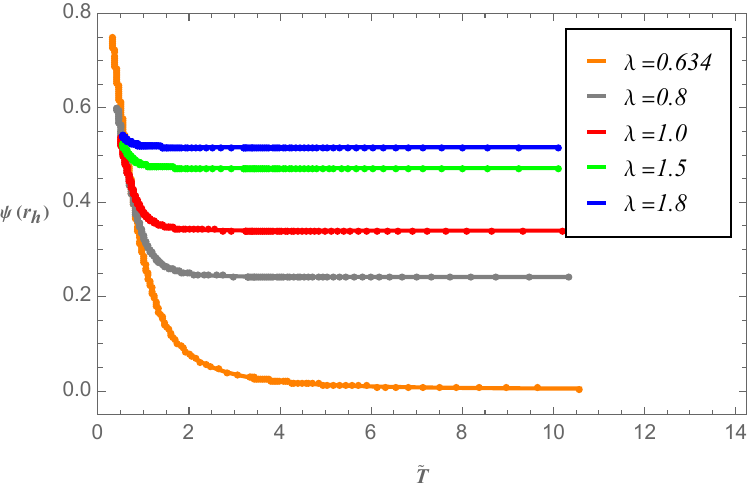}
    \caption{Scalar condensation at the horizon $\psi(r_h)$ as the function of temperature with different Gauss-Bonnet coupling $\lambda$ in the situation of holographic superconductor (left) and spontaneous scalarization (right), respectively.}
    \label{fig=holo_sca_phase}
\end{figure}

Numerically, one may ascertain that the two types of hairy black holes are indeed distinct when they are further apart in the parameter space.
This can be demonstrated, for instance, by examining the scalar condensation near the horizon $\psi(r_h)$ as a function of the temperature for different Gauss-Bonnet coupling constants.
The results are presented in Fig.~\ref{fig=holo_sca_phase}.
In the left panel, for a holographic superconductor, the results indicate how the system undergoes a phase transition.
The scalar condensation merges as the system cools down and reaches the transition temperature represented by the solid black curve in Fig.~\ref{fig=phase-diagram}.
It rapidly grows as the temperature further decreases. 
However, as the coupling approaches the critical value $\lambda_c$, the transition temperature increases and eventually diverges.
The dependence of the condensation on temperature becomes less drastic, and the overall shape of the curve approaches that of the spontaneous scalarization, as shown on the right panel of Fig.~\ref{fig=holo_sca_phase}.
The latter features distinctive patterns in two temperature regimes.
At elevated temperatures, the scalar condensation converges to a constant magnitude that increases with increasing Gauss-Bonnet coupling.
As temperature decreases, a surge of scalar condensation is observed.
This behavior differs from spontaneous scalarization in the absence of an electromagnetic field due to an interplay between the two instabilities. 

However, the above analysis, and the phase diagram Fig.~\ref{fig=phase-diagram}, are primarily based on the coupling constant $\lambda$, which does not possess a straightforward thermodynamic interpretation.
It is still not entirely clear whether there is a phase transition from a superconductor phase to the spontaneous scalarization one.
In particular, it seems not straightforward to elaborate a thermodynamic process through which a superconducting black hole transits into a scalarized one by raising or decreasing the temperature.
Regarding the AdS/CFT dictionary, it is more meaningful to present the results in terms of intensive thermodynamic quantities such as temperature and chemical potential.
Such an approach is carried out in the following subsection.
Moreover, we evaluate the system's free energy and elaborate further on the properties of the phase transition.

\subsection{Transitions among the black holes}\label{sec=HBH_transition}

\begin{figure}[thbp]
    \centering
    \includegraphics[width=0.48\linewidth]{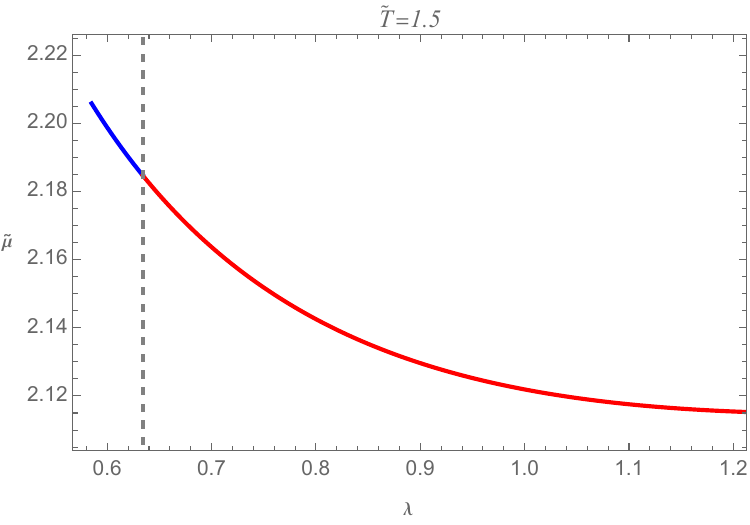}
    \includegraphics[width=0.48\linewidth]{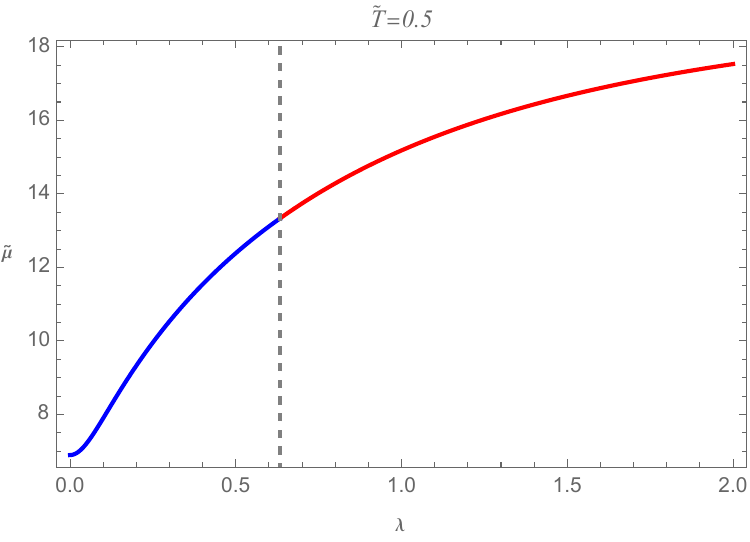}
    \includegraphics[width=0.48\linewidth]{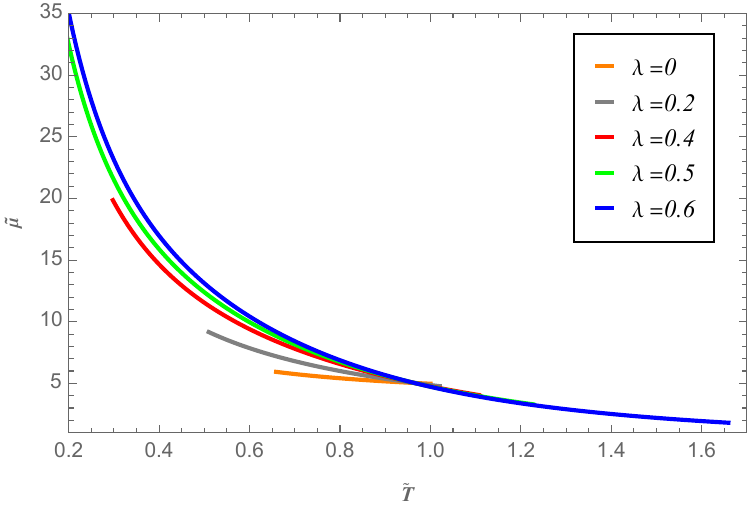}
    \includegraphics[width=0.48\linewidth]{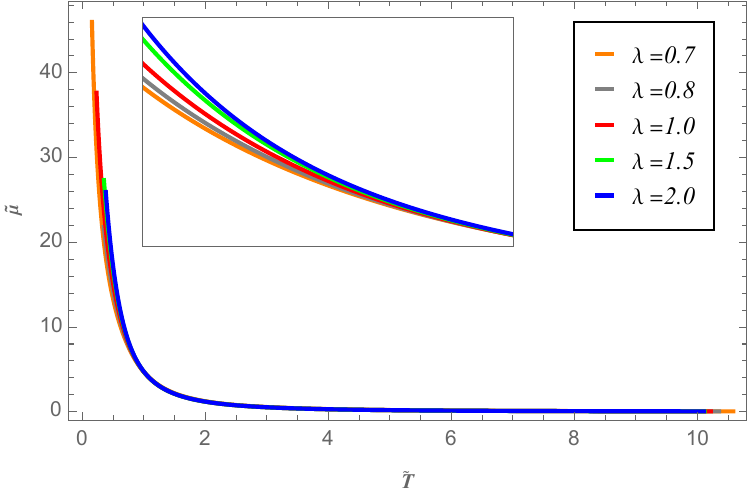}
    \caption{The top row shows the chemical potential as a function of the Gauss-Bonnet coupling in the high temperature (top-left) and low temperature (top-right) regions, where the blue curves represent the holographic superconductors while the red curves describe the spontaneous scalarization, and the vertical gray dashed line is the critical coupling $\lambda_c$.
    The bottom row shows the chemical potential as a function of temperature for different Gauss-Bonnet couplings.}
    \label{fig=mu_lambda}
\end{figure}

The phase diagram presented in Fig.~\ref{fig=phase-diagram} is closely related to the numerical employed scheme, which is standard in the literature. 
However, to explore the underlying phase transition between the two hairy black holes, it is meaningful to show the phase diagram in terms of thermodynamic variables such as temperature and chemical potential.
Using the AdS/CFT dictionary, one can reiterate the phase diagram shown in Fig.~\ref{fig=phase-diagram} in terms of intensive quantities, namely, temperature and chemical potential.
The results are presented in Fig.~\ref{fig=phase-diagram2}.
It is noted that the chemical potential does not remain constant when the system evolves along a vertical line with given $\lambda$, as shown in the bottom row of Fig.~\ref{fig=mu_lambda}.
When comparing Fig.~\ref{fig=phase-diagram2} against Fig.~\ref{fig=phase-diagram}, one observes a few intriguing features of the phase structure of the system.

The solid black curve bridges the transition between the holographically superconducting black hole and a hairless one, as shown in the left panel of Fig.~\ref{fig=phase-diagram2}.
The latter corresponds to the region of elevated temperature and lower chemical potential.
A dashed black curve indicates the transition between the holographic superconductor and spontaneous scalarization, which will be elaborated further by evaluating the free energy.
In this region, one may consider the following process with given chemical potential for an initially thermalized bald black hole with elevated temperature.
As the temperature gradually decreases, the scalar field will condensate and form a holographic supercondutor via a second-order phase transition by tranversing the solid black curve.
Subsequently, as the temperature further decreases, the system transits into a scalarized black hole through a first-order transition by crossing the dashed black curve.
However, the above qualitative properties regarding the phase division does not change as one goes to higher-temperature region.
In other words, even though the solid black curve and dashed black one asymptotically approach each other in Fig.~\ref{fig=phase-diagram} at the high-temperature limit, the two-phase transitions do not actually intersect in the phase space presented in terms of temperature and chemical potential.
Moreover, the spontaneous scalarization phase does not extend and occupies the remainder of the phase space. 
It can be verified numerically that the temperature and chemical potential remain finite even at the limit $\lambda\to +\infty$.
Also, a background hairless RN black hole spans the entire parameter space.

The middle panel of Fig.~\ref{fig=phase-diagram2} shows a similar transition, but the positions of the two phases are exchanged.
It is also noted that the superconducting black hole cannot transit to a hairless counterpart in this region.
This is because the solid black curve terminates at $\tilde{T}=1$ in Fig.~\ref{fig=phase-diagram} as such a transition is confined in the region $T>T_c$.

The results in the left and middle panels imply that the two phases successively flip to the other side of the transition curve in a narrow region of the phase diagram, as shown in the right panel of Fig.~\ref{fig=phase-diagram2}.
Such an intriguing phenomenon can be attributed to the non-monotonic behavior of chemical potential as a function of Gauss-Bonnet coupling at different temperatures.
At an elevated temperature related to the left panel of Fig.~\ref{fig=phase-diagram2}, the chemical potential decreases as the coupling constant increases.
Conversely, at a lower temperature corresponding to the middle panel of Fig.~\ref{fig=phase-diagram2}, the chemical potential increases monotonically with increasing coupling.
The above results are explicitly shown in the top-left and top-right panels of Fig.~\ref{fig=mu_lambda}.

To provide a more comprehensive analysis of the transition between the holographic superconductor and spontaneous scalarization phases, we evaluate the on-shell Gibbs free energy of the system by following~\cite{Basak:2015vma, Ghorai:2016tvk}, which reads
\begin{equation}
F= \frac{F_{\Omega}}{V_2}=-\frac{1}{2}\mu\rho-\psi_1\psi_2+\int^1_0dz\bigg(\frac{q^2\phi(z)^2\psi(z)^2}{z^2g(z)}-\frac{\lambda^2\mathcal{R}^2_{GB}}{2z^4}\psi(z)^4\bigg)~, \label{free-energy}
\end{equation}
with $z=\frac{r_h}{r}$ here and elaborate on the Gibbs condition.

\begin{figure}[thbp]
    \centering
    \includegraphics[width=0.325\linewidth]{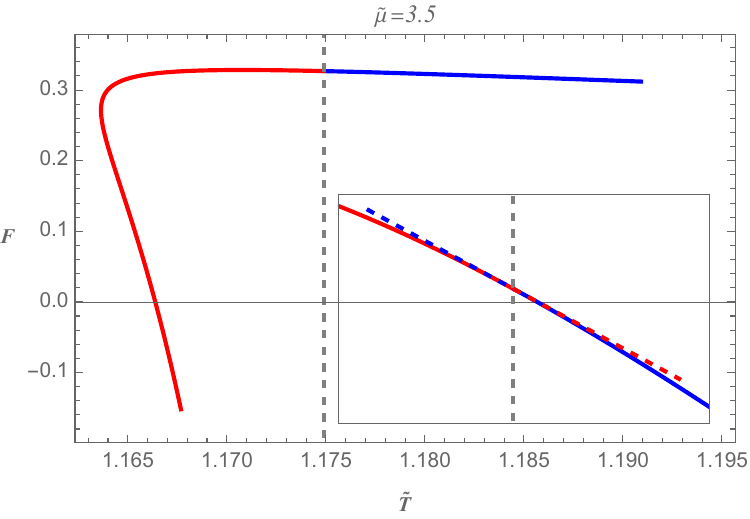}
    \includegraphics[width=0.325\linewidth]{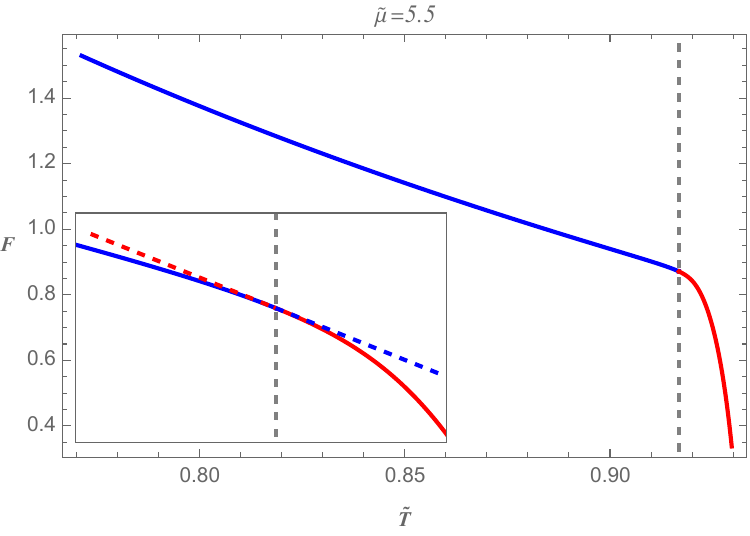}
    \includegraphics[width=0.325\linewidth]{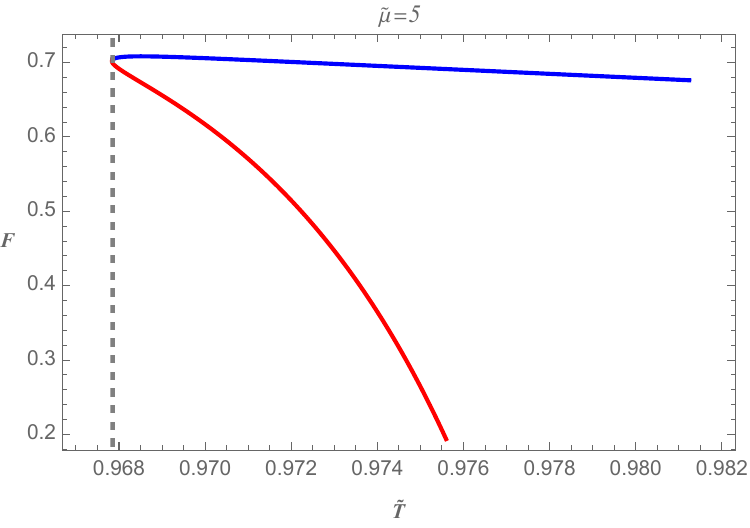}
    \includegraphics[width=0.325\linewidth]{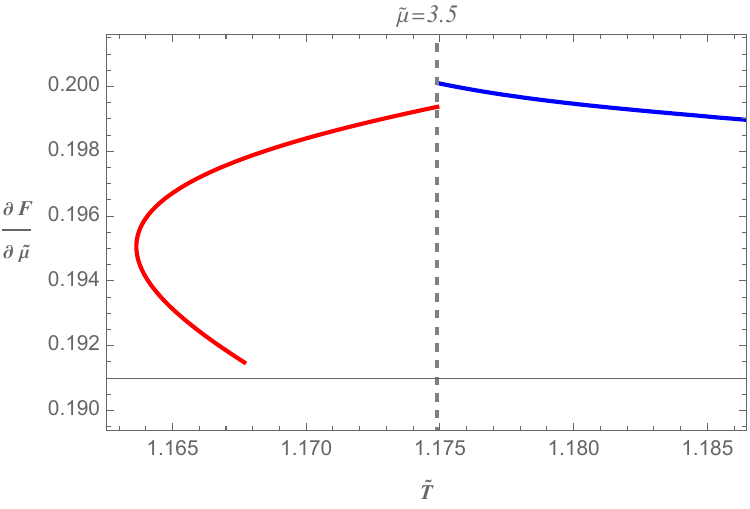}
    \includegraphics[width=0.325\linewidth]{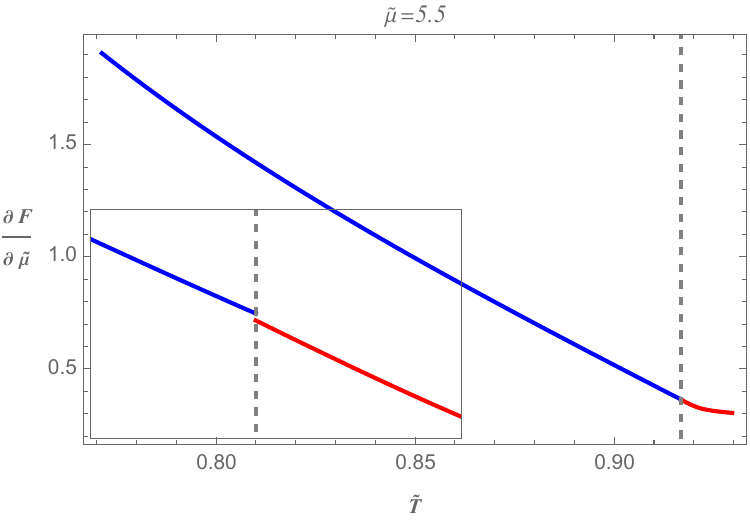}
    \includegraphics[width=0.325\linewidth]{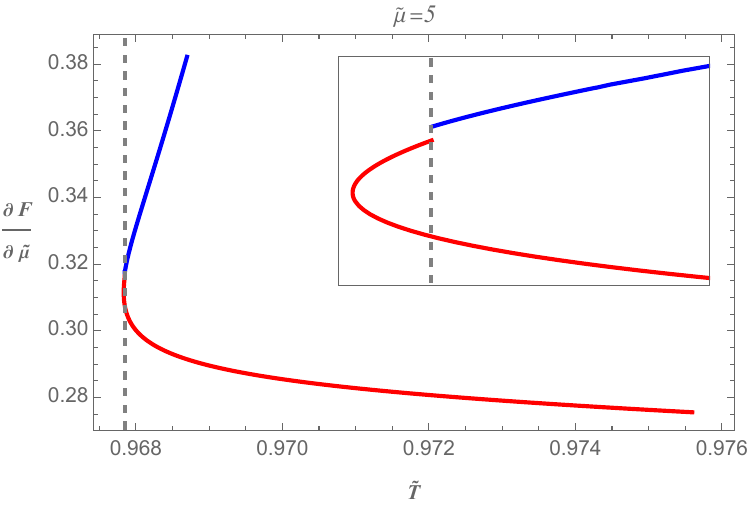}
    \caption{The free energy and its first-order derivative w.r.t. the chemical potential as functions of temperature for different chemical potentials.
    The blue curves represent the holographic superconductors while the red curves describe the spontaneous scalarization, and the vertical gray dashed line indicates the transition point.}
    \label{fig=free_energy}
\end{figure}

\begin{figure}[thbp]
    \centering
    \includegraphics[width=0.325\linewidth]{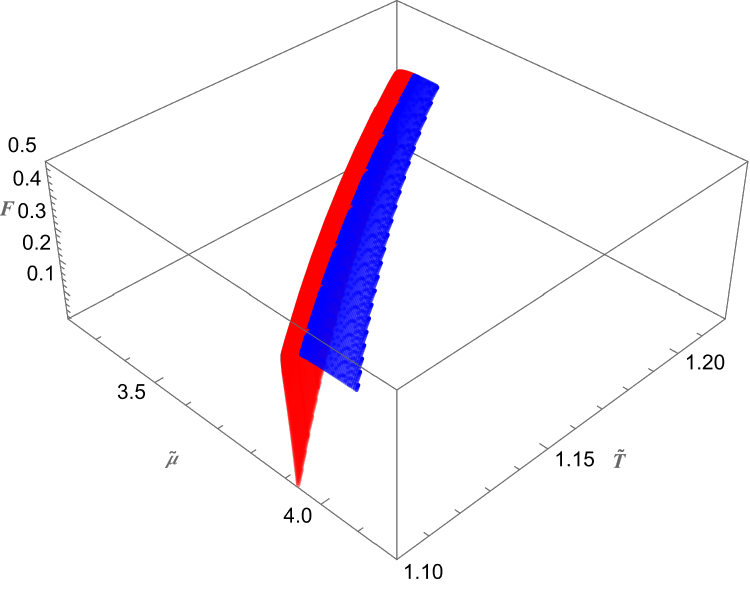}
    \includegraphics[width=0.325\linewidth]{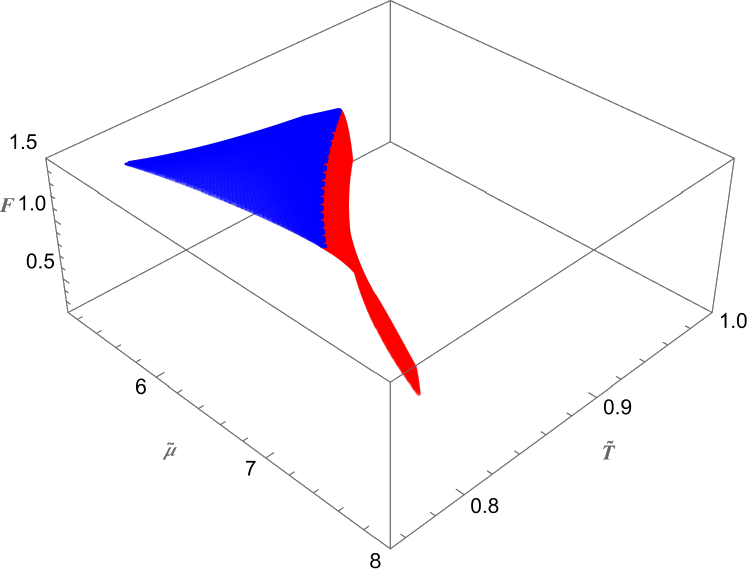}
    \includegraphics[width=0.325\linewidth]{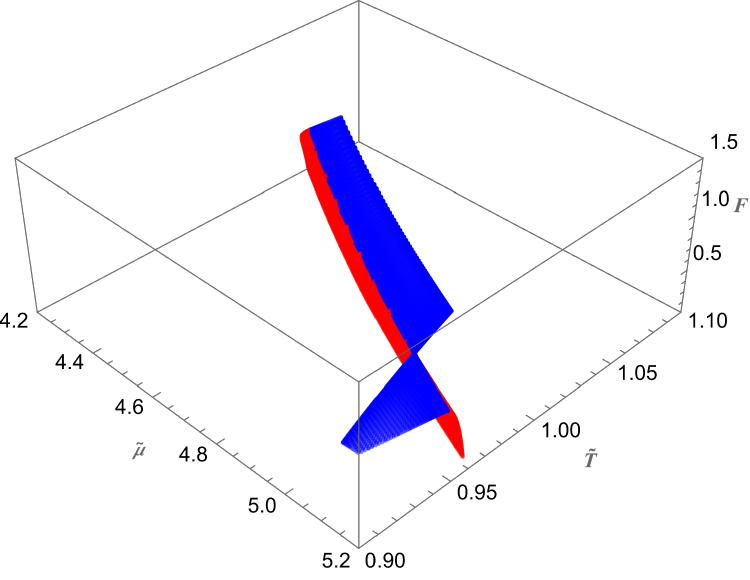}
    \caption{The system's free energy as a function of chemical potential and temperature.}
    \label{fig=free_energy_3d}
\end{figure}

The results are shown in Figs.~\ref{fig=free_energy} and~\ref{fig=free_energy_3d}.
The three plots on the first row of Fig.~\ref{fig=free_energy} demonstrate the Gibbs conditions for, respectively, typical scenarios corresponding to the left, middle, and right panels of Fig.~\ref{fig=phase-diagram2}.
Gibbs condition dictates that a phase transition occurs with equalized temperature, pressure, and chemical potential.
When there is competition between the two mechanisms for hairy black holes, the surviving state corresponds to the one with less free energy.
The magnified sections of the plots indicate that the transition is of first order for the first two plots since the free energy shot above the other phase after the transition point.
The latter is also confirmed by explicitly showing that the first-order derivatives of the free energy are discontinuous, as given in the second row of Fig.~\ref{fig=free_energy}.
We therefore confirm that one can elaborate a thermodynamic process through which a superconducting black hole transits into a scalarized one by raising or decreasing the temperature.
The purple region in the bottom panel of Fig.~\ref{fig=phase-diagram2} indicates an intriguing scenario.
There, the two phases directly compete in the purple region of the right panel of Fig.~\ref{fig=phase-diagram2}.
According to the free energy evaluated and presented in the last plot on the first row of Fig.~\ref{fig=free_energy}, a scalarized black hole possesses a smaller free energy and, therefore, is favorable.

In Fig.~\ref{fig=free_energy_3d}, we present the free energy landscape from a three-dimensional perspective.
It illustrates how the transition takes place in a more intuitive fashion.
The two free-energy surfaces intersect at a dashed black curve.
When comparing the left and middle panels, one observes that the two phases switch their positions with respect to the transition curve.
This occurs near the critical temperature $\tilde{T} \gtrsim 1$.
There, the free-energy surface of the spontaneous scalarization is found to swing from one side to the other through the vertical direction, while the superconducting phase's surface varies moderately, and the slope remains essentially unchanged.
For the superconducting phase, the phase transition to bald black hole does not occur for $\tilde{T}<1$, and subsequently, the transition curve represented by the solid black curves ends at $\tilde{T}=1$.
The latter region $\tilde{T}<1$ bounded by $\lambda=0$ also flips to the other side of the transition curve.
These characteristics lead to the exchange of the two phases, as discussed in Fig.~\ref{fig=free_energy}.
Also, as mentioned above, for the region where the two phases coexist, a scalarized black hole always possesses a smaller free energy than the superconducting one and, therefore, is more favorable.

\subsection{Instability analysis using effective potential}\label{sec=instab}

In this subsection, we complement our analysis by studying the tachyonic instability of the effective potentials.
On the one hand, for a holographic superconductor, it is understood that the coupling between the scalar and Maxwell field in asymptotic AdS spacetime leads to tachyonic instability~\cite{Hartnoll:2008kx}. 
On the other hand, the coupling between the scalar field and the Gauss-Bonnet curvature also gives rise to tachyonic instability~\cite{Silva:2017uqg, Silva:2018qhn, Antoniou:2021zoy}, and it has been argued that the origin of the instability is Gregory-Laflamme type~\cite{Myung:2018iyq}. 
This section delves into the effective potential Eq.~\eqref{eq=potential} of the scalar perturbations regarding the underlying instabilities.
In literature, the tachyonic instability is primarily attributed to a negative effective mass extracted from examining the master equation of scalar perturbations.
For a holographic superconductor, a negative effective scalar mass arises from the non-vanishing the Maxwell field (electrostatic potential) in AdS spacetime.
For spontaneous scalarization, this is due to the contribution coming from the non-vanishing coupling between the scalar and higher curvature term.
In what follows, the numerical results of the effective potential Eq.~\eqref{eq=potential} are obtained by solving the system of equations, Eqs.~\eqref{eq=phi} and~\eqref{eq=psi}, for given $\lambda$, $q$, and $\phi'(r_h)=0.5$ instead of $\psi(r_h)$.

\begin{figure}[thbp]
    \centering
    \includegraphics[width=0.48\linewidth]{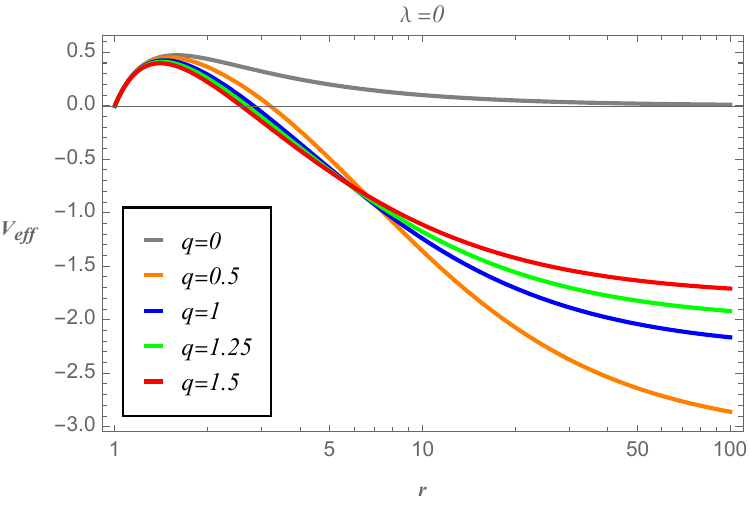}
    \includegraphics[width=0.48\linewidth]{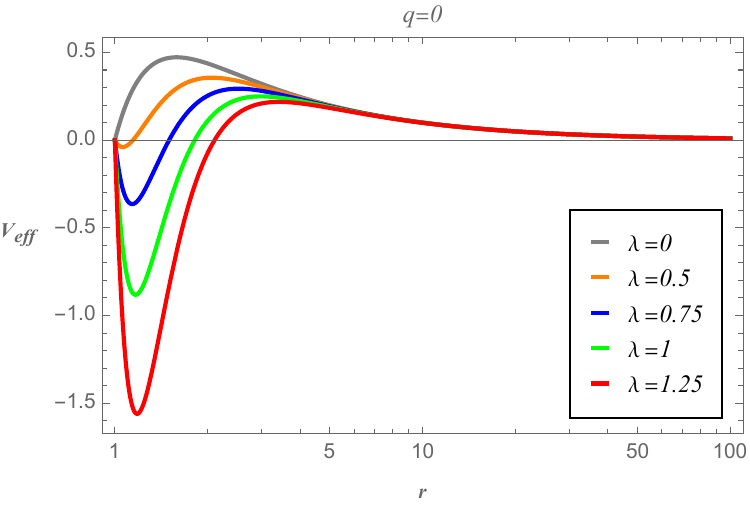}
    \includegraphics[width=0.48\linewidth]{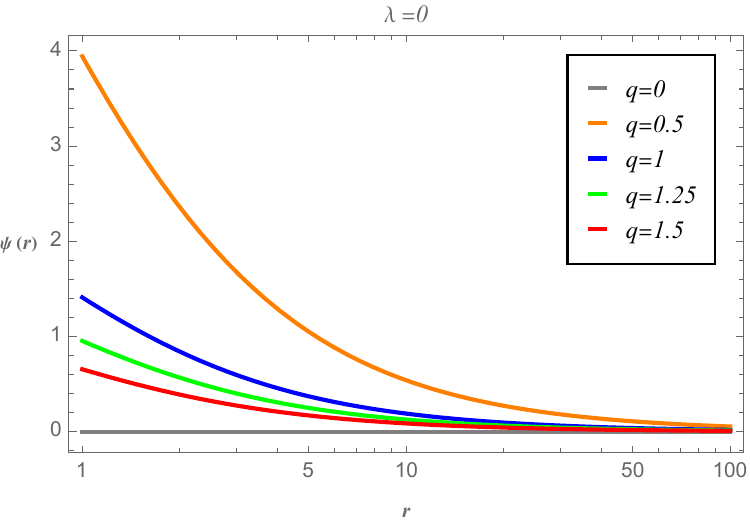}
    \includegraphics[width=0.48\linewidth]{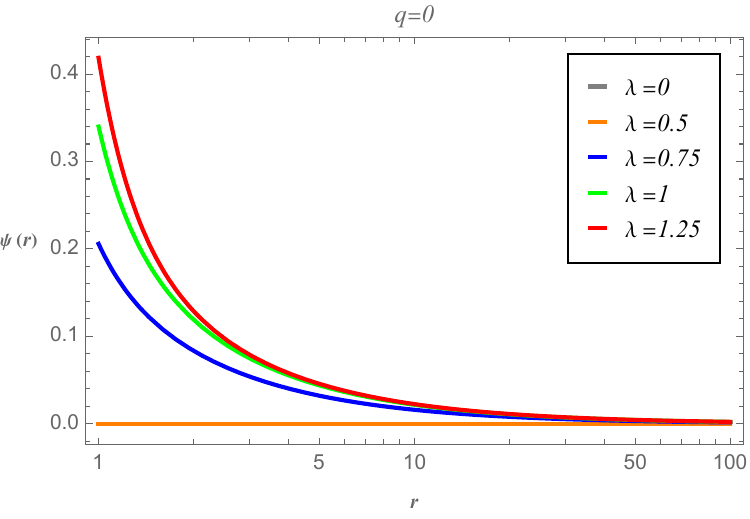}
    \caption{The top row shows the calculated effective potentials for different model parameters.
    Top-left: The effective potential for the holographic superconductors with $\lambda=0$, evaluated for different charges $q$.
    Top-right: The effective potential for the spontaneously scalarized black hole with $q=0$, evaluated for different values of $\lambda$.
    The bottom row shows the corresponding profiles of the scalar field $\psi(r)$.}
    \label{fig=holo_poten}
\end{figure}

In Fig.~\ref{fig=holo_poten}, we present the effective potentials and the corresponding profiles of the scalar field, evaluated for various metric parameters.
As pointed out in~\cite{Myung:2018iyq}, the effective potential of tachyonic instability in AdS spacetime is featured by a positive barrier near the event horizon that smoothly converges to a given value at infinity.
In the top-left panel of Fig.~\ref{fig=holo_poten}, the resulting effective potentials of a pure holographic superconductor agree with such a behavior. 
In the present model, this can be achieved by assuming $\lambda=0$.
We note that the effective potential asymptotically approaches $V_\mathrm{eff}\to -q^2\phi(r)^2$ at spatial infinity for the given metric parameters.
The different curves in the plot are obtained by taking different values for the charge $q$.
A positive potential barrier is formed near the event horizon for all the cases. 
An increase of the charge $q$ causes the potential barrier to dissipate less rapidly as the radial coordinate increases.
The negative values of the effective potential are attributed to the condensation of the Maxwell field, giving rise to tachyonic instability in scalar perturbations.
Also, a more significant coupling between the scalar and Maxwell fields leads to a higher transition temperature for holographic superconductivity~\cite{Hartnoll:2008kx}.

Conversely, a pure spontaneously scalarized black hole metric can be obtained in the present model by taking $q=0$. 
The corresponding effective potentials are evaluated and shown in the top-right panel of Fig.~\ref{fig=holo_poten}, where one varies the Gauss-Bonnet coupling $\lambda$.
One observes that the obtained effective potential's main feature differs from that of a holographic superconductor. 
In particular, a potential well is formed near the event horizon, and the depth of this well gradually increases as the Gauss-Bonnet coupling $\lambda$ grows.
Therefore, the potential well close to the horizon caused by the Gauss-Bonnet coupling is primarily understood to cause the tachyonic instability.
Such a characteristic of the effective potential has been extensively discussed in the literature of spontaneous scalarization~\cite{Silva:2017uqg, Blazquez-Salcedo:2018jnn}, which was attributed to the Gregory-Laflamme instability in Ref.~\cite{Myung:2018iyq}.
The above discussions further confirm the distinct nature of the holographic superconductivity and spontaneous scalarization phases, which are substantially triggered by entirely distinct mechanisms. 
Besides the thermodynamic quantities, such as the free energy, the difference is also manifested by the effective potential. 

\begin{figure}[thbp]
    \centering
    \includegraphics[width=0.48\linewidth]{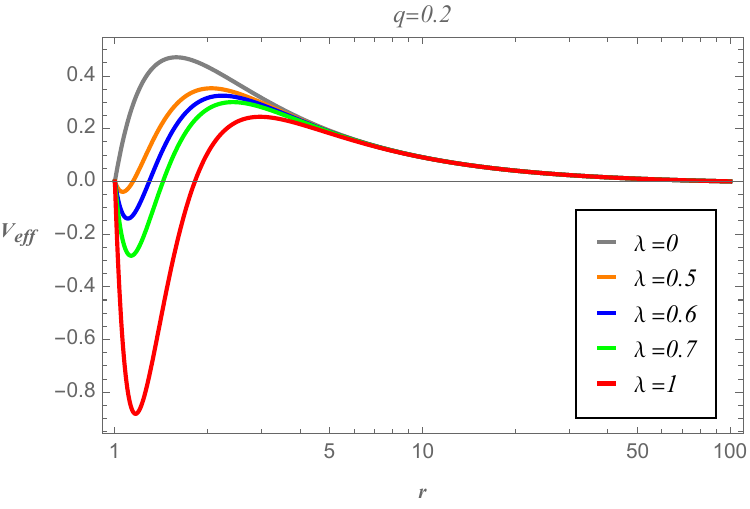}
    \includegraphics[width=0.48\linewidth]{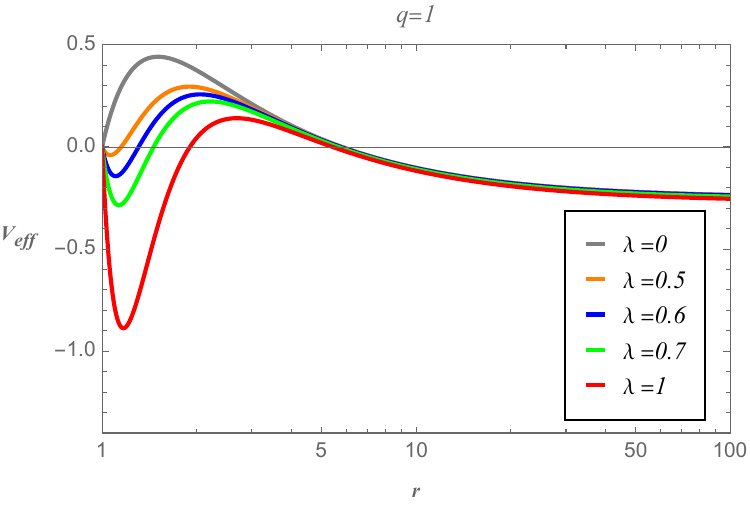}
    \caption{The calculated effective potential with different couplings between the scalar and Maxwell fields.
    The calculations are carried out for different charges $q$.
    Left: The effective potential for a smaller coupling $q=0.2$.
    Right: The effective potential for a more significant coupling $q=1$.}
    \label{fig=mix_poten_lambda}
\end{figure}

The effective potentials are further explored in Fig.~\ref{fig=mix_poten_lambda} for different values of the charge $q$.
It is found that a more significant Gauss-Bonnet coupling $\lambda$ deepens the potential well near the horizon without affecting its behavior at infinity. 
Also, a more significant value of $q$ causes the value of the effective potential to become more negative at infinity while not influencing the potential well near the event horizon. 
Therefore, one concludes that both mechanisms play a role in the present model rather independently.
Subsequently, it is interesting to explore the properties of the hairy black hole in the region where the two mechanisms compete.

\begin{figure}[thbp]
    \centering
    \includegraphics[width=0.48\linewidth]{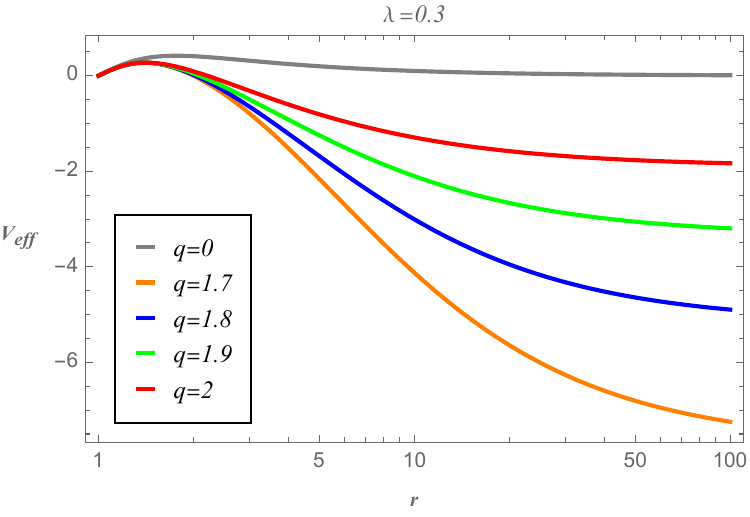}
    \includegraphics[width=0.48\linewidth]{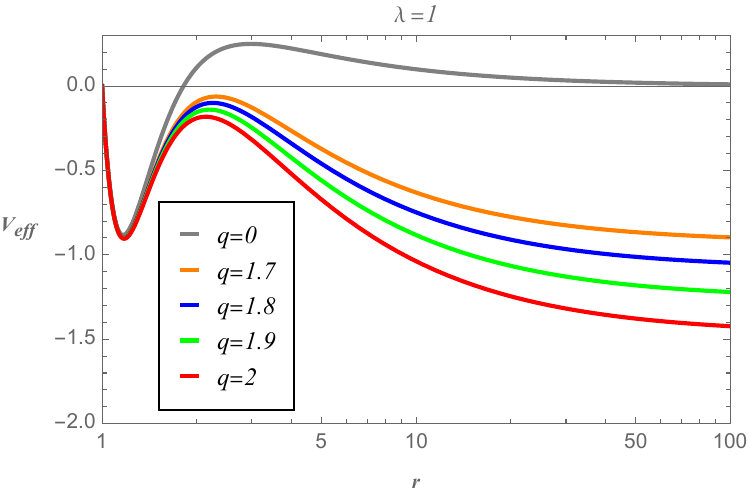}
    \caption{The calculated effective potential near the transition curve between a holographic superconductor and spontaneous scalarization.
    The calculations are carried out for different charges $q$.
    Left: The effective potential for the holographic superconductor with $\lambda \lesssim \lambda_c$.
    Right: The effective potential for the spontaneously scalarized black hole with $\lambda \gtrsim \lambda_c$.}
    \label{fig=mix_poten_q}
\end{figure}

In the preceding subsections, we elaborated on a scenario where both phases are present and explored the transition between them.
It is, therefore, also interesting to study the effective potentials in the phase transition region.
This analysis is presented in Fig.~\ref{fig=mix_poten_q}, showing the calculated effective potential near the transition curve.
The calculations are carried out for different charges $q$.
In the left panel of Fig.~\ref{fig=mix_poten_q}, the behavior of the effective potential closely resembles that of a pure holographic superconductor. 
In other words, the obtained effective potential clearly demonstrates the physical mechanism that gives birth to a holographic superconductor.
On the other hand, in the right panel of Fig.~\ref{fig=mix_poten_q}, where the Gauss-Bonnet coupling surpasses the critical value $\lambda>\lambda_c$, the effective potential forms a well outside the event horizon. 
While the presence of the potential well does not affect its asymptotical form at large radial coordinates, it suffices for the tachyonic instability to trigger the scalarization.
It is observed that varying $q$ does not alter the depth of the potential well near the horizon, as such instability is irrelevant to the Maxwell field. 
Notably, when comparing the left and right panels of Fig.~\ref{fig=mix_poten_q}, one observes that the resulting effective potentials are pretty similar, which further gives rise to similar profiles as discussed in Fig.~\ref{fig=holo_sca_phase}.  
To distinguish the two phases, we have to resort to explicit calculations of the free energy.

\section{Further discussions and concluding remarks}\label{sec=conclusion}

We employed a model that unites both mechanisms based on existing studies on holographic superconductors and spontaneous scalarization~\cite{Guo:2020sdu,Zou:2023inv,Chen:2022vag,Bao:2021wfu,Brihaye:2019dck}.
The system's phase space is scrutinized numerically by developing a high-precision shooting method. 
The spontaneous scalarization phase always prevails as long as the Gauss-Bonnet coupling exceeds the critical value $\lambda>\lambda_{c}$. 
The phase diagram is shown in Fig.~\ref{fig=phase-diagram} in terms of $(\tilde{T},\lambda)$. 
Although the transition from a bald black hole to a superconductor is well-established and is of second-order phase transition, it is somehow obscure to us whether there is a well-defined phase transition from the superconductor to a scalarized black hole, given that both the radial profile and effective potential between the two phases are somehow indistinguishable.
Also, it needs to be clarified if the system transits from one phase to another by simply raising or decreasing the temperature.
This seeming ambiguity is further explored by examining the phase diagram in terms of temperature and chemical potential $(\tilde{T},\tilde{\mu})$, as shown in Fig.~\ref{fig=phase-diagram2}.
By explicitly evaluating the Gibbs free energy and its derivatives, it is shown that such a phase transition is well-defined and of first order. 
Moreover, the phase diagram indicates a non-trivial feature as the two phases flip over to the other side along the transition curve.
These results reinforced that the spontaneous scalarization and holographic superconductor do not transit smoothly between one another and are potentially induced by different instabilities. 

Last but not least, we mention a few potentially related topics to the present work.
In our calculations, we have only considered the fundamental modes.
The scalar perturbations have been shown~\cite{Silva:2017uqg} to possess further excited states in the decoupling limit.
Also, spontaneous scalarization induced by other matter sources, such as the Einstein-Maxwell-scalar theory~\cite{Guo:2021zed, Chen:2023eru}, is also a worthy possibility. 
We have focused on a minimal toy model that comprises both phases, and a more generic background black hole metric further spans the model's parameter space while reflecting a more realistic scenario.
Regarding thermal properties, we elaborated on the complexity during the transition between these two phases, and it is interesting to probe such properties at zero temperature~\cite{Marrani:2022hva}. 
The relation between the quasinormal modes and instability has been extensively explored in the literature.
Such analysis in the context of phase transition is also relevant to the present model. 
Given that the present study primarily adopts the probe limit, it is imperative to consider numerical calculations involving backreaction. 
Relevant studies regarding holographic superconductor have been carried out~\cite{Hartnoll:2008kx}, and a generalization of the present scheme will provide further insights beyond the linearized theory.
Moreover, a study of the dual phase of scalarization was performed in~\cite{Brihaye:2019dck}. 
Further generalization is an intriguing direction.
Applications of the present findings to the strongly coupled quantum system and condensed matter physics might be beneficial. 



\begin{acknowledgments}
This research is supported by the financial support from Brazilian agencies Funda\c{c}\~ao de Amparo \`a Pesquisa do Estado de S\~ao Paulo (FAPESP), Funda\c{c}\~ao de Amparo \`a Pesquisa do Estado do Rio de Janeiro (FAPERJ), Conselho Nacional de Desenvolvimento Cient\'{\i}fico e Tecnol\'ogico (CNPq), and Coordena\c{c}\~ao de Aperfei\c{c}oamento de Pessoal de N\'ivel Superior (CAPES).

\end{acknowledgments}

\bibliographystyle{jhep}
\bibliography{holosca_Guo, references_qian}

\providecommand{\href}[2]{#2}\begingroup\raggedright\begin{thebibliography}{10}

\bibitem{book-blackhole-Frolov}
V.~P. Frolov and I.~D. Novikov, \emph{Black Hole Physics: Basic Concepts and
  New Developments}.
\newblock Kluwer Academic, 1998,
  \href{http://dx.doi.org/10.1007/978-94-011-5139-9}{10.1007/978-94-011-5139-9}.

\bibitem{Berti:2015itd}
E.~Berti et~al., \emph{{Testing General Relativity with Present and Future
  Astrophysical Observations}},
  \href{http://dx.doi.org/10.1088/0264-9381/32/24/243001}{\emph{Class. Quant.
  Grav.} {\bf 32} (2015) 243001}, [\href{http://arxiv.org/abs/1501.07274}{{\tt
  1501.07274}}].

\bibitem{LIGOScientific:2016vlm}
{\scshape LIGO Scientific, Virgo} collaboration, B.~P. Abbott et~al.,
  \emph{{Properties of the Binary Black Hole Merger GW150914}},
  \href{http://dx.doi.org/10.1103/PhysRevLett.116.241102}{\emph{Phys. Rev.
  Lett.} {\bf 116} (2016) 241102}, [\href{http://arxiv.org/abs/1602.03840}{{\tt
  1602.03840}}].

\bibitem{LIGOScientific:2018mvr}
{\scshape LIGO Scientific, Virgo} collaboration, B.~P. Abbott et~al.,
  \emph{{GWTC-1: A Gravitational-Wave Transient Catalog of Compact Binary
  Mergers Observed by LIGO and Virgo during the First and Second Observing
  Runs}}, \href{http://dx.doi.org/10.1103/PhysRevX.9.031040}{\emph{Phys. Rev.
  X} {\bf 9} (2019) 031040}, [\href{http://arxiv.org/abs/1811.12907}{{\tt
  1811.12907}}].

\bibitem{LIGOScientific:2019fpa}
{\scshape LIGO Scientific, Virgo} collaboration, B.~P. Abbott et~al.,
  \emph{{Tests of General Relativity with the Binary Black Hole Signals from
  the LIGO-Virgo Catalog GWTC-1}},
  \href{http://dx.doi.org/10.1103/PhysRevD.100.104036}{\emph{Phys. Rev. D} {\bf
  100} (2019) 104036}, [\href{http://arxiv.org/abs/1903.04467}{{\tt
  1903.04467}}].

\bibitem{LIGOScientific:2021djp}
{\scshape LIGO Scientific, VIRGO, KAGRA} collaboration, R.~Abbott et~al.,
  \emph{{GWTC-3: Compact Binary Coalescences Observed by LIGO and Virgo During
  the Second Part of the Third Observing Run}},
  \href{http://arxiv.org/abs/2111.03606}{{\tt 2111.03606}}.

\bibitem{Barack:2018yly}
L.~Barack et~al., \emph{{Black holes, gravitational waves and fundamental
  physics: a roadmap}},
  \href{http://dx.doi.org/10.1088/1361-6382/ab0587}{\emph{Class. Quant. Grav.}
  {\bf 36} (2019) 143001}, [\href{http://arxiv.org/abs/1806.05195}{{\tt
  1806.05195}}].

\bibitem{Berry:2019wgg}
C.~P.~L. Berry, S.~A. Hughes, C.~F. Sopuerta, A.~J.~K. Chua, A.~Heffernan,
  K.~Holley-Bockelmann et~al., \emph{{The unique potential of extreme
  mass-ratio inspirals for gravitational-wave astronomy}},
  \href{http://arxiv.org/abs/1903.03686}{{\tt 1903.03686}}.

\bibitem{Babak:2017tow}
S.~Babak, J.~Gair, A.~Sesana, E.~Barausse, C.~F. Sopuerta, C.~P.~L. Berry
  et~al., \emph{{Science with the space-based interferometer LISA. V: Extreme
  mass-ratio inspirals}},
  \href{http://dx.doi.org/10.1103/PhysRevD.95.103012}{\emph{Phys. Rev. D} {\bf
  95} (2017) 103012}, [\href{http://arxiv.org/abs/1703.09722}{{\tt
  1703.09722}}].

\bibitem{Ruffini:1971bza}
R.~Ruffini and J.~A. Wheeler, \emph{{Introducing the black hole}},
  \href{http://dx.doi.org/10.1063/1.3022513}{\emph{Phys. Today} {\bf 24} (1971)
  30}.

\bibitem{Chrusciel:2012jk}
P.~T. Chrusciel, J.~Lopes~Costa and M.~Heusler, \emph{{Stationary Black Holes:
  Uniqueness and Beyond}},
  \href{http://dx.doi.org/10.12942/lrr-2012-7}{\emph{Living Rev. Rel.} {\bf 15}
  (2012) 7}, [\href{http://arxiv.org/abs/1205.6112}{{\tt 1205.6112}}].

\bibitem{Carter:1971zc}
B.~Carter, \emph{{Axisymmetric Black Hole Has Only Two Degrees of Freedom}},
  \href{http://dx.doi.org/10.1103/PhysRevLett.26.331}{\emph{Phys. Rev. Lett.}
  {\bf 26} (1971) 331--333}.

\bibitem{Herdeiro:2015waa}
C.~A.~R. Herdeiro and E.~Radu, \emph{{Asymptotically flat black holes with
  scalar hair: a review}},
  \href{http://dx.doi.org/10.1142/S0218271815420146}{\emph{Int. J. Mod. Phys.
  D} {\bf 24} (2015) 1542014}, [\href{http://arxiv.org/abs/1504.08209}{{\tt
  1504.08209}}].

\bibitem{Sotiriou:2015pka}
T.~P. Sotiriou, \emph{{Black Holes and Scalar Fields}},
  \href{http://dx.doi.org/10.1088/0264-9381/32/21/214002}{\emph{Class. Quant.
  Grav.} {\bf 32} (2015) 214002}, [\href{http://arxiv.org/abs/1505.00248}{{\tt
  1505.00248}}].

\bibitem{Herdeiro:2014goa}
C.~A.~R. Herdeiro and E.~Radu, \emph{{Kerr black holes with scalar hair}},
  \href{http://dx.doi.org/10.1103/PhysRevLett.112.221101}{\emph{Phys. Rev.
  Lett.} {\bf 112} (2014) 221101}, [\href{http://arxiv.org/abs/1403.2757}{{\tt
  1403.2757}}].

\bibitem{Gubser:2008px}
S.~S. Gubser, \emph{{Breaking an Abelian gauge symmetry near a black hole
  horizon}}, \href{http://dx.doi.org/10.1103/PhysRevD.78.065034}{\emph{Phys.
  Rev. D} {\bf 78} (2008) 065034}, [\href{http://arxiv.org/abs/0801.2977}{{\tt
  0801.2977}}].

\bibitem{Hartnoll:2008kx}
S.~A. Hartnoll, C.~P. Herzog and G.~T. Horowitz, \emph{{Holographic
  Superconductors}},
  \href{http://dx.doi.org/10.1088/1126-6708/2008/12/015}{\emph{JHEP} {\bf 12}
  (2008) 015}, [\href{http://arxiv.org/abs/0810.1563}{{\tt 0810.1563}}].

\bibitem{Sudarsky:2002mk}
D.~Sudarsky and J.~A. Gonzalez, \emph{{On black hole scalar hair in
  asymptotically anti-de Sitter space-times}},
  \href{http://dx.doi.org/10.1103/PhysRevD.67.024038}{\emph{Phys. Rev. D} {\bf
  67} (2003) 024038}, [\href{http://arxiv.org/abs/gr-qc/0207069}{{\tt
  gr-qc/0207069}}].

\bibitem{Doneva:2017bvd}
D.~D. Doneva and S.~S. Yazadjiev, \emph{{New Gauss-Bonnet Black Holes with
  Curvature-Induced Scalarization in Extended Scalar-Tensor Theories}},
  \href{http://dx.doi.org/10.1103/PhysRevLett.120.131103}{\emph{Phys. Rev.
  Lett.} {\bf 120} (2018) 131103}, [\href{http://arxiv.org/abs/1711.01187}{{\tt
  1711.01187}}].

\bibitem{Silva:2017uqg}
H.~O. Silva, J.~Sakstein, L.~Gualtieri, T.~P. Sotiriou and E.~Berti,
  \emph{{Spontaneous scalarization of black holes and compact stars from a
  Gauss-Bonnet coupling}},
  \href{http://dx.doi.org/10.1103/PhysRevLett.120.131104}{\emph{Phys. Rev.
  Lett.} {\bf 120} (2018) 131104}, [\href{http://arxiv.org/abs/1711.02080}{{\tt
  1711.02080}}].

\bibitem{Antoniou:2017acq}
G.~Antoniou, A.~Bakopoulos and P.~Kanti, \emph{{Evasion of No-Hair Theorems and
  Novel Black-Hole Solutions in Gauss-Bonnet Theories}},
  \href{http://dx.doi.org/10.1103/PhysRevLett.120.131102}{\emph{Phys. Rev.
  Lett.} {\bf 120} (2018) 131102}, [\href{http://arxiv.org/abs/1711.03390}{{\tt
  1711.03390}}].

\bibitem{Herdeiro:2018wub}
C.~A.~R. Herdeiro, E.~Radu, N.~Sanchis-Gual and J.~A. Font, \emph{{Spontaneous
  Scalarization of Charged Black Holes}},
  \href{http://dx.doi.org/10.1103/PhysRevLett.121.101102}{\emph{Phys. Rev.
  Lett.} {\bf 121} (2018) 101102}, [\href{http://arxiv.org/abs/1806.05190}{{\tt
  1806.05190}}].

\bibitem{Brihaye:2018bgc}
Y.~Brihaye, C.~Herdeiro and E.~Radu, \emph{{The scalarised Schwarzschild-NUT
  spacetime}},
  \href{http://dx.doi.org/10.1016/j.physletb.2018.11.022}{\emph{Phys. Lett. B}
  {\bf 788} (2019) 295--301}, [\href{http://arxiv.org/abs/1810.09560}{{\tt
  1810.09560}}].

\bibitem{Herdeiro:2019yjy}
C.~A.~R. Herdeiro and E.~Radu, \emph{{Black hole scalarization from the
  breakdown of scale invariance}},
  \href{http://dx.doi.org/10.1103/PhysRevD.99.084039}{\emph{Phys. Rev. D} {\bf
  99} (2019) 084039}, [\href{http://arxiv.org/abs/1901.02953}{{\tt
  1901.02953}}].

\bibitem{Dima:2020yac}
A.~Dima, E.~Barausse, N.~Franchini and T.~P. Sotiriou, \emph{{Spin-induced
  black hole spontaneous scalarization}},
  \href{http://dx.doi.org/10.1103/PhysRevLett.125.231101}{\emph{Phys. Rev.
  Lett.} {\bf 125} (2020) 231101}, [\href{http://arxiv.org/abs/2006.03095}{{\tt
  2006.03095}}].

\bibitem{Herdeiro:2020wei}
C.~A.~R. Herdeiro, E.~Radu, H.~O. Silva, T.~P. Sotiriou and N.~Yunes,
  \emph{{Spin-induced scalarized black holes}},
  \href{http://dx.doi.org/10.1103/PhysRevLett.126.011103}{\emph{Phys. Rev.
  Lett.} {\bf 126} (2021) 011103}, [\href{http://arxiv.org/abs/2009.03904}{{\tt
  2009.03904}}].

\bibitem{Berti:2020kgk}
E.~Berti, L.~G. Collodel, B.~Kleihaus and J.~Kunz, \emph{{Spin-induced
  black-hole scalarization in Einstein-scalar-Gauss-Bonnet theory}},
  \href{http://dx.doi.org/10.1103/PhysRevLett.126.011104}{\emph{Phys. Rev.
  Lett.} {\bf 126} (2021) 011104}, [\href{http://arxiv.org/abs/2009.03905}{{\tt
  2009.03905}}].

\bibitem{Doneva:2023oww}
D.~D. Doneva, L.~Arest\'e~Sal\'o, K.~Clough, P.~Figueras and S.~S. Yazadjiev,
  \emph{{Testing the limits of scalar-Gauss-Bonnet gravity through nonlinear
  evolutions of spin-induced scalarization}},
  \href{http://dx.doi.org/10.1103/PhysRevD.108.084017}{\emph{Phys. Rev. D} {\bf
  108} (2023) 084017}, [\href{http://arxiv.org/abs/2307.06474}{{\tt
  2307.06474}}].

\bibitem{Doneva:2021tvn}
D.~D. Doneva and S.~S. Yazadjiev, \emph{{Beyond the spontaneous scalarization:
  New fully nonlinear mechanism for the formation of scalarized black holes and
  its dynamical development}},
  \href{http://dx.doi.org/10.1103/PhysRevD.105.L041502}{\emph{Phys. Rev. D}
  {\bf 105} (2022) L041502}, [\href{http://arxiv.org/abs/2107.01738}{{\tt
  2107.01738}}].

\bibitem{Lai:2023gwe}
M.-Y. Lai, D.-C. Zou, R.-H. Yue and Y.~S. Myung, \emph{{Nonlinearly scalarized
  rotating black holes in Einstein-scalar-Gauss-Bonnet theory}},
  \href{http://dx.doi.org/10.1103/PhysRevD.108.084007}{\emph{Phys. Rev. D} {\bf
  108} (2023) 084007}, [\href{http://arxiv.org/abs/2304.08012}{{\tt
  2304.08012}}].

\bibitem{Doneva:2022yqu}
D.~D. Doneva, L.~G. Collodel and S.~S. Yazadjiev, \emph{{Spontaneous nonlinear
  scalarization of Kerr black holes}},
  \href{http://dx.doi.org/10.1103/PhysRevD.106.104027}{\emph{Phys. Rev. D} {\bf
  106} (2022) 104027}, [\href{http://arxiv.org/abs/2208.02077}{{\tt
  2208.02077}}].

\bibitem{Pombo:2023lxg}
A.~M. Pombo and D.~D. Doneva, \emph{{Effects of mass and self-interaction on
  nonlinear scalarization of scalar-Gauss-Bonnet black holes}},
  \href{http://arxiv.org/abs/2310.08638}{{\tt 2310.08638}}.

\bibitem{Silva:2020omi}
H.~O. Silva, H.~Witek, M.~Elley and N.~Yunes, \emph{{Dynamical Descalarization
  in Binary Black Hole Mergers}},
  \href{http://dx.doi.org/10.1103/PhysRevLett.127.031101}{\emph{Phys. Rev.
  Lett.} {\bf 127} (2021) 031101}, [\href{http://arxiv.org/abs/2012.10436}{{\tt
  2012.10436}}].

\bibitem{Kuan:2021lol}
H.-J. Kuan, D.~D. Doneva and S.~S. Yazadjiev, \emph{{Dynamical Formation of
  Scalarized Black Holes and Neutron Stars through Stellar Core Collapse}},
  \href{http://dx.doi.org/10.1103/PhysRevLett.127.161103}{\emph{Phys. Rev.
  Lett.} {\bf 127} (2021) 161103}, [\href{http://arxiv.org/abs/2103.11999}{{\tt
  2103.11999}}].

\bibitem{Zhang:2021etr}
C.-Y. Zhang, P.~Liu, Y.~Liu, C.~Niu and B.~Wang, \emph{{Dynamical charged black
  hole spontaneous scalarization in anti\textendash{}de Sitter spacetimes}},
  \href{http://dx.doi.org/10.1103/PhysRevD.104.084089}{\emph{Phys. Rev. D} {\bf
  104} (2021) 084089}, [\href{http://arxiv.org/abs/2103.13599}{{\tt
  2103.13599}}].

\bibitem{Zhang:2021ybj}
C.-Y. Zhang, P.~Liu, Y.~Liu, C.~Niu and B.~Wang, \emph{{Dynamical scalarization
  in Einstein-Maxwell-dilaton theory}},
  \href{http://dx.doi.org/10.1103/PhysRevD.105.024073}{\emph{Phys. Rev. D} {\bf
  105} (2022) 024073}, [\href{http://arxiv.org/abs/2111.10744}{{\tt
  2111.10744}}].

\bibitem{Zhang:2021nnn}
C.-Y. Zhang, Q.~Chen, Y.~Liu, W.-K. Luo, Y.~Tian and B.~Wang, \emph{{Critical
  Phenomena in Dynamical Scalarization of Charged Black Holes}},
  \href{http://dx.doi.org/10.1103/PhysRevLett.128.161105}{\emph{Phys. Rev.
  Lett.} {\bf 128} (2022) 161105}, [\href{http://arxiv.org/abs/2112.07455}{{\tt
  2112.07455}}].

\bibitem{Liu:2022fxy}
Y.~Liu, C.-Y. Zhang, Q.~Chen, Z.~Cao, Y.~Tian and B.~Wang, \emph{{Critical
  scalarization and descalarization of black holes in a generalized
  scalar-tensor theory}},
  \href{http://dx.doi.org/10.1007/s11433-023-2160-1}{\emph{Sci. China Phys.
  Mech. Astron.} {\bf 66} (2023) 100412},
  [\href{http://arxiv.org/abs/2208.07548}{{\tt 2208.07548}}].

\bibitem{Bakopoulos:2018nui}
A.~Bakopoulos, G.~Antoniou and P.~Kanti, \emph{{Novel Black-Hole Solutions in
  Einstein-Scalar-Gauss-Bonnet Theories with a Cosmological Constant}},
  \href{http://dx.doi.org/10.1103/PhysRevD.99.064003}{\emph{Phys. Rev. D} {\bf
  99} (2019) 064003}, [\href{http://arxiv.org/abs/1812.06941}{{\tt
  1812.06941}}].

\bibitem{Brihaye:2019gla}
Y.~Brihaye, C.~Herdeiro and E.~Radu, \emph{{Black Hole Spontaneous
  Scalarisation with a Positive Cosmological Constant}},
  \href{http://dx.doi.org/10.1016/j.physletb.2020.135269}{\emph{Phys. Lett. B}
  {\bf 802} (2020) 135269}, [\href{http://arxiv.org/abs/1910.05286}{{\tt
  1910.05286}}].

\bibitem{Brihaye:2019dck}
Y.~Brihaye, B.~Hartmann, N.~P. Aprile and J.~Urrestilla, \emph{{Scalarization
  of asymptotically anti\textendash{}de Sitter black holes with applications to
  holographic phase transitions}},
  \href{http://dx.doi.org/10.1103/PhysRevD.101.124016}{\emph{Phys. Rev. D} {\bf
  101} (2020) 124016}, [\href{http://arxiv.org/abs/1911.01950}{{\tt
  1911.01950}}].

\bibitem{Guo:2021zed}
G.~Guo, P.~Wang, H.~Wu and H.~Yang, \emph{{Scalarized
  Einstein\textendash{}Maxwell-scalar black holes in anti-de Sitter
  spacetime}},
  \href{http://dx.doi.org/10.1140/epjc/s10052-021-09614-7}{\emph{Eur. Phys. J.
  C} {\bf 81} (2021) 864}, [\href{http://arxiv.org/abs/2102.04015}{{\tt
  2102.04015}}].

\bibitem{Promsiri:2023yda}
C.~Promsiri, T.~Tangphati, E.~Hirunsirisawat and S.~Ponglertsakul,
  \emph{{Scalarization of planar anti\textendash{}de Sitter charged black holes
  in Einstein-Maxwell-scalar theory}},
  \href{http://dx.doi.org/10.1103/PhysRevD.108.024015}{\emph{Phys. Rev. D} {\bf
  108} (2023) 024015}, [\href{http://arxiv.org/abs/2302.04654}{{\tt
  2302.04654}}].

\bibitem{Zou:2023inv}
D.-C. Zou, B.~Meng, M.~Zhang, S.-Y. Li, M.-Y. Lai and Y.~S. Myung,
  \emph{{Analytical approximate solutions for scalarized AdS black holes}},
  \href{http://dx.doi.org/10.3390/universe9010026}{\emph{Universe} {\bf 9}
  (2023) 26}, [\href{http://arxiv.org/abs/2301.04784}{{\tt 2301.04784}}].

\bibitem{Marrani:2022hva}
A.~Marrani, O.~Miskovic and P.~Q. Leon, \emph{{Spontaneous scalarization in
  (A)dS gravity at zero temperature}},
  \href{http://dx.doi.org/10.1007/JHEP07(2022)100}{\emph{JHEP} {\bf 07} (2022)
  100}, [\href{http://arxiv.org/abs/2203.14388}{{\tt 2203.14388}}].

\bibitem{Doneva:2022ewd}
D.~D. Doneva, F.~M. Ramazano\u{g}lu, H.~O. Silva, T.~P. Sotiriou and S.~S.
  Yazadjiev, \emph{{Scalarization}},
  \href{http://arxiv.org/abs/2211.01766}{{\tt 2211.01766}}.

\bibitem{Blazquez-Salcedo:2018jnn}
J.~L. Bl\'azquez-Salcedo, D.~D. Doneva, J.~Kunz and S.~S. Yazadjiev,
  \emph{{Radial perturbations of the scalarized Einstein-Gauss-Bonnet black
  holes}}, \href{http://dx.doi.org/10.1103/PhysRevD.98.084011}{\emph{Phys. Rev.
  D} {\bf 98} (2018) 084011}, [\href{http://arxiv.org/abs/1805.05755}{{\tt
  1805.05755}}].

\bibitem{Myung:2018iyq}
Y.~S. Myung and D.-C. Zou, \emph{{Gregory-Laflamme instability of black hole in
  Einstein-scalar-Gauss-Bonnet theories}},
  \href{http://dx.doi.org/10.1103/PhysRevD.98.024030}{\emph{Phys. Rev. D} {\bf
  98} (2018) 024030}, [\href{http://arxiv.org/abs/1805.05023}{{\tt
  1805.05023}}].

\bibitem{Silva:2018qhn}
H.~O. Silva, C.~F.~B. Macedo, T.~P. Sotiriou, L.~Gualtieri, J.~Sakstein and
  E.~Berti, \emph{{Stability of scalarized black hole solutions in
  scalar-Gauss-Bonnet gravity}},
  \href{http://dx.doi.org/10.1103/PhysRevD.99.064011}{\emph{Phys. Rev. D} {\bf
  99} (2019) 064011}, [\href{http://arxiv.org/abs/1812.05590}{{\tt
  1812.05590}}].

\bibitem{Antoniou:2021zoy}
G.~Antoniou, A.~Leh\'ebel, G.~Ventagli and T.~P. Sotiriou, \emph{{Black hole
  scalarization with Gauss-Bonnet and Ricci scalar couplings}},
  \href{http://dx.doi.org/10.1103/PhysRevD.104.044002}{\emph{Phys. Rev. D} {\bf
  104} (2021) 044002}, [\href{http://arxiv.org/abs/2105.04479}{{\tt
  2105.04479}}].

\bibitem{buell1995potentials}
W.~F. Buell and B.~Shadwick, \emph{Potentials and bound states},
  {\emph{American Journal of Physics} {\bf 63} (1995) 256--258}.

\bibitem{Brigante:2008gz}
M.~Brigante, H.~Liu, R.~C. Myers, S.~Shenker and S.~Yaida, \emph{{The Viscosity
  Bound and Causality Violation}},
  \href{http://dx.doi.org/10.1103/PhysRevLett.100.191601}{\emph{Phys. Rev.
  Lett.} {\bf 100} (2008) 191601}, [\href{http://arxiv.org/abs/0802.3318}{{\tt
  0802.3318}}].

\bibitem{Brigante:2007nu}
M.~Brigante, H.~Liu, R.~C. Myers, S.~Shenker and S.~Yaida, \emph{{Viscosity
  Bound Violation in Higher Derivative Gravity}},
  \href{http://dx.doi.org/10.1103/PhysRevD.77.126006}{\emph{Phys. Rev. D} {\bf
  77} (2008) 126006}, [\href{http://arxiv.org/abs/0712.0805}{{\tt 0712.0805}}].

\bibitem{Buchel:2009tt}
A.~Buchel and R.~C. Myers, \emph{{Causality of Holographic Hydrodynamics}},
  \href{http://dx.doi.org/10.1088/1126-6708/2009/08/016}{\emph{JHEP} {\bf 08}
  (2009) 016}, [\href{http://arxiv.org/abs/0906.2922}{{\tt 0906.2922}}].

\bibitem{Hofman:2008ar}
D.~M. Hofman and J.~Maldacena, \emph{{Conformal collider physics: Energy and
  charge correlations}},
  \href{http://dx.doi.org/10.1088/1126-6708/2008/05/012}{\emph{JHEP} {\bf 05}
  (2008) 012}, [\href{http://arxiv.org/abs/0803.1467}{{\tt 0803.1467}}].

\bibitem{Hofman:2009ug}
D.~M. Hofman, \emph{{Higher Derivative Gravity, Causality and Positivity of
  Energy in a UV complete QFT}},
  \href{http://dx.doi.org/10.1016/j.nuclphysb.2009.08.001}{\emph{Nucl. Phys. B}
  {\bf 823} (2009) 174--194}, [\href{http://arxiv.org/abs/0907.1625}{{\tt
  0907.1625}}].

\bibitem{Camanho:2009vw}
X.~O. Camanho and J.~D. Edelstein, \emph{{Causality constraints in AdS/CFT from
  conformal collider physics and Gauss-Bonnet gravity}},
  \href{http://dx.doi.org/10.1007/JHEP04(2010)007}{\emph{JHEP} {\bf 04} (2010)
  007}, [\href{http://arxiv.org/abs/0911.3160}{{\tt 0911.3160}}].

\bibitem{Gregory:1993vy}
R.~Gregory and R.~Laflamme, \emph{{Black strings and p-branes are unstable}},
  \href{http://dx.doi.org/10.1103/PhysRevLett.70.2837}{\emph{Phys. Rev. Lett.}
  {\bf 70} (1993) 2837--2840}, [\href{http://arxiv.org/abs/hep-th/9301052}{{\tt
  hep-th/9301052}}].

\bibitem{Hartnoll:2008vx}
S.~A. Hartnoll, C.~P. Herzog and G.~T. Horowitz, \emph{{Building a Holographic
  Superconductor}},
  \href{http://dx.doi.org/10.1103/PhysRevLett.101.031601}{\emph{Phys. Rev.
  Lett.} {\bf 101} (2008) 031601}, [\href{http://arxiv.org/abs/0803.3295}{{\tt
  0803.3295}}].

\bibitem{Guo:2020sdu}
H.~Guo, S.~Kiorpelidi, X.-M. Kuang, E.~Papantonopoulos, B.~Wang and J.-P. Wu,
  \emph{{Spontaneous holographic scalarization of black holes in
  Einstein-scalar-Gauss-Bonnet theories}},
  \href{http://dx.doi.org/10.1103/PhysRevD.102.084029}{\emph{Phys. Rev. D} {\bf
  102} (2020) 084029}, [\href{http://arxiv.org/abs/2006.10659}{{\tt
  2006.10659}}].

\bibitem{Guo:2020zqm}
H.~Guo, X.-M. Kuang, E.~Papantonopoulos and B.~Wang, \emph{{Horizon curvature
  and spacetime structure influences on black hole scalarization}},
  \href{http://dx.doi.org/10.1140/epjc/s10052-021-09630-7}{\emph{Eur. Phys. J.
  C} {\bf 81} (2021) 842}, [\href{http://arxiv.org/abs/2012.11844}{{\tt
  2012.11844}}].

\bibitem{Franco:2009yz}
S.~Franco, A.~Garcia-Garcia and D.~Rodriguez-Gomez, \emph{{A General class of
  holographic superconductors}},
  \href{http://dx.doi.org/10.1007/JHEP04(2010)092}{\emph{JHEP} {\bf 04} (2010)
  092}, [\href{http://arxiv.org/abs/0906.1214}{{\tt 0906.1214}}].

\bibitem{Bao:2021wfu}
Y.~Bao, H.~Guo and X.-M. Kuang, \emph{{Excited states of holographic
  superconductor with scalar field coupled to Gauss-Bonnet invariance}},
  \href{http://dx.doi.org/10.1016/j.physletb.2021.136646}{\emph{Phys. Lett. B}
  {\bf 822} (2021) 136646}.

\bibitem{Pan:2021jii}
J.~Pan, X.~Qiao, D.~Wang, Q.~Pan, Z.-Y. Nie and J.~Jing, \emph{{Holographic
  superconductors in 4D Einstein-Gauss-Bonnet gravity with backreactions}},
  \href{http://dx.doi.org/10.1016/j.physletb.2021.136755}{\emph{Phys. Lett. B}
  {\bf 823} (2021) 136755}, [\href{http://arxiv.org/abs/2109.02207}{{\tt
  2109.02207}}].

\bibitem{Basak:2015vma}
S.~Basak, P.~Chaturvedi, P.~Nandi and G.~Sengupta, \emph{{Thermodynamic
  geometry of holographic superconductors}},
  \href{http://dx.doi.org/10.1016/j.physletb.2015.12.061}{\emph{Phys. Lett. B}
  {\bf 753} (2016) 493--499}, [\href{http://arxiv.org/abs/1509.00826}{{\tt
  1509.00826}}].

\bibitem{Ghorai:2016tvk}
D.~Ghorai and S.~Gangopadhyay, \emph{{Holographic free energy and thermodynamic
  geometry}},
  \href{http://dx.doi.org/10.1140/epjc/s10052-016-4555-1}{\emph{Eur. Phys. J.
  C} {\bf 76} (2016) 702}, [\href{http://arxiv.org/abs/1607.05187}{{\tt
  1607.05187}}].

\bibitem{Chen:2022vag}
Q.~Chen, Z.~Ning, Y.~Tian, B.~Wang and C.-Y. Zhang, \emph{{Descalarization by
  quenching charged hairy black hole in asymptotically AdS spacetime}},
  \href{http://dx.doi.org/10.1007/JHEP01(2023)062}{\emph{JHEP} {\bf 01} (2023)
  062}, [\href{http://arxiv.org/abs/2210.14539}{{\tt 2210.14539}}].

\bibitem{Chen:2023eru}
Q.~Chen, Z.~Ning, Y.~Tian, B.~Wang and C.-Y. Zhang, \emph{{Nonlinear dynamics
  of hot, cold, and bald Einstein-Maxwell-scalar black holes in AdS
  spacetime}}, \href{http://dx.doi.org/10.1103/PhysRevD.108.084016}{\emph{Phys.
  Rev. D} {\bf 108} (2023) 084016},
  [\href{http://arxiv.org/abs/2307.03060}{{\tt 2307.03060}}].

\end{thebibliography}\endgroup
\end{document}